\documentclass[onecolumn,showpacs,12pt,superscriptaddress,prb]{revtex4}
\usepackage{amsmath}
\usepackage{latexsym}
\usepackage{graphicx}
\usepackage{amsfonts}
\usepackage{amssymb}
\usepackage{appendix}
\usepackage{graphics}
\usepackage{mathrsfs,lipsum}
\numberwithin{equation}{section}

\begin{document}

\title{Numerical Analysis of the Helmholtz Green's Function for Scalar Wave Propagation Through a Nano-hole on a Plasmonic Layer}
\author{D\'{e}sir\'{e} Miessein }
\affiliation{Department of Physics and Engineering Physics,
Fordham University, Bronx, New York 10458, USA}
\affiliation{Department of Physics and Engineering Physics,
Stevens Institute of Technology, Hoboken, New Jersey 07030, USA}
\author{Norman J. M. Horing}
\affiliation{Department of Physics and Engineering Physics,
Stevens Institute of Technology, Hoboken, New Jersey 07030, USA}
\author{Godfrey Gumbs}
\affiliation{Department of Physics and Astronomy,\\
Hunter College of the City University of New York, New York, NY 10065, USA}
\affiliation{Donostia International Physics Center (DIPC),
P de Manuel Lardizabal, 4, 20018 San Sebastian, Basque Country, Spain}
\author{Harry Lenzing}
\affiliation{Department of Physics and Engineering Physics,
Stevens Institute of Technology, Hoboken, New Jersey 07030, USA}
\date{\today}
\begin{abstract}
\begin{center}
    Abstract
\end{center}
A detailed numerical study of the Helmholtz Green's function for the description of scalar wave propagation through a nano-hole on a plasmonic layer
is presented here.  In conjunction with this, we briefly review the analytic formulation taking the nano-hole radius as the smallest length parameter
of the system.  Figures exhibiting the numerical results for this Green's function in various ranges of the transmission region are presented.
\end{abstract}
\maketitle
\def\thesection{\arabic{section}}
\def\thesubsection{\arabic{subsection}}
\section{Introduction}
\label{sec1}
The transmission properties of a scalar field propagating through a nano-hole in a two-dimensional (2D) plasmonic layer have been analyzed
using a Green's function technique in conjunction with an integral equation formulation [1,2,3,4].
The nano-hole is taken to lie on a plasmonic sheet (located on the plane $ z = 0 $ embedded in a three-dimensional (3D) bulk host
medium with background dielectric constant $\varepsilon_{b}^{(3D)}$). In Section II of this paper, we briefly review in some detail the analytic determination of the scalar Helmholtz Green's function with the presence of the layer in which a two-dimensional plasma is embedded.
Section III reviews the scalar Helmholtz Green's function solution for the 2D plasmonic layer embedded in a 3D host medium with the presence of a nano-hole aperture in the subwavelength regime.  The results of our thorough numerical analysis of the perforated layer Helmholtz Green's function are discussed in Section IV with illustrative figures showing results in the near, middle and far field zones of the transmission region.  Finally, conclusions are summarized in Section V.
\newpage
\numberwithin{subsection}{section}
\numberwithin{subsubsection}{subsection}
\numberwithin{equation}{subsection}

\section{Green's Function Solution for Full 2D Plasmonic Layer Embedded in a 3D Bulk Host Medium}
\label{sec2}
\subsection{Integral Equation for the Scalar Green's Function and Solution}
\numberwithin{figure}{subsection}
\begin{figure}[h]
\centering
\includegraphics[width=17cm,height=7cm]{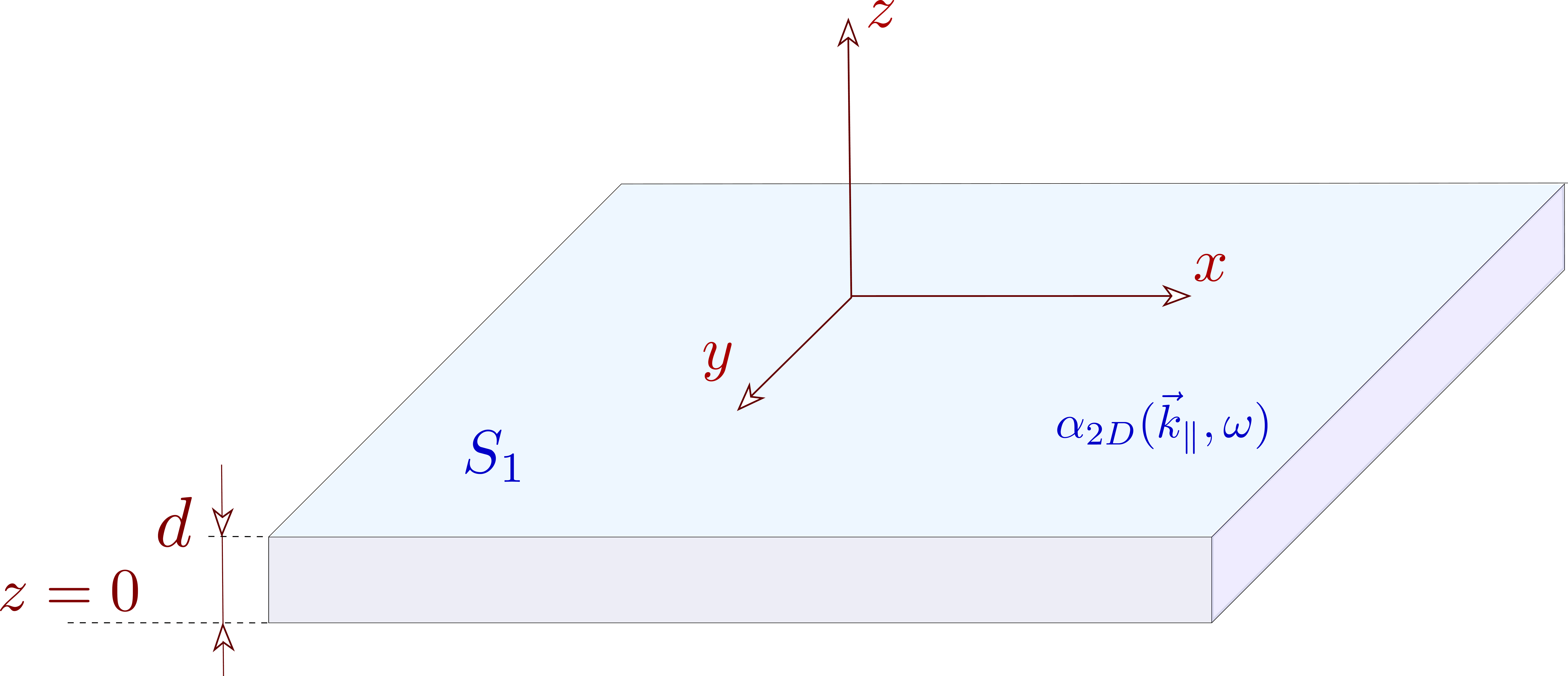}
\caption{(Color online)\ Schematic illustration of a two dimensional plasmonic layer $ S_{1}$ of thickness
$ d $ embedded at $ z=0 $ in a three dimensional bulk meduim with dielectric
constant $ \varepsilon_{b}^{(3D)}$.}
\label{FIG211R}
\qquad
\end{figure}
\par
 We consider a two dimensional plasmonic layer $ S_{1}$ to have a dynamic nonlocal $2D$ polarizability
  $ \alpha^{(2D)}(\vec{k}_{\parallel},\omega)$, located on the plane
$ z = 0 $, embedded in a three dimensional bulk host medium with
background dielectric constant $\varepsilon_{b}^{(3D)}$
(Fig.\ref{FIG211R}).  The associated Helmholtz Green's function including the two-dimensional plasmonic sheet,
$ G_{fs}$ \textit{without} a nano-hole, satisfies the integro-differential equation (position/frequency
representation)[1,2,3,4]
\begin{eqnarray}\label{2.1}
 \left(-\vec{\nabla}^{2}-\frac{\omega^{2}}{c^2}\varepsilon_{b}^{(3D)}\right)
G_{fs}(\vec{r},\vec{r}^{\prime};\omega)
- \,\,\frac{\omega^{2}}{c^2}
\int{d^{3}\vec{r}^{'}}\alpha^{(2D)}_{fs}(\vec{r},\vec{r}^{''};\omega)G_{fs}(\vec{r}^{''},\vec{r}^{'};\omega)=\delta^{(3)}( \vec{r}-\vec{r}^{'}).
\nonumber\\
\end{eqnarray}
The polarizability $ \alpha^{(2D)}_{fs}$ of the full 2D plasmonic layer has the form
\begin{equation}\label{2.2}
\alpha^{(2D)}_{fs}(\vec{r},\vec{r}^{'};\omega)=\alpha^{(2D)}_{fs}(\vec{r}_{\parallel},\vec{r}^{'}_{\parallel};\omega)\,d\, \delta(z) \delta(z')
\end{equation}
where $ d $ is the thickness of the plasmonic sheet, $ \vec{r}=(\vec{r}_{\parallel}; z)$ and $ \alpha^{(2D)}_{fs}(\vec{r}_{\parallel},\vec{r}^{'}_{\parallel};\omega)$ is the 2D plasmonic polarizability of the 2D sheet; $\delta(z) $  is the Dirac delta function needed to confine the polarizability onto the plane of the 2D layer at $ z = 0 $.

\medskip
\par
To solve Eq.\ (\ref{2.1}), we employ the bulk Helmholtz Green's function [5]

\begin{equation}\label{2.3I0}
\left(-\vec{\nabla}^{2}-\frac{\omega^{2}}{c^2}\varepsilon_{b}^{(3D)}\right)G_{3D}(\vec{r},\vec{r}^{'};\omega)
= \delta^{3D}(\vec{r}-\vec{r}^{'}).
\end{equation}
After performing the 2D spatial Fourier transform of $ G_{3D}$ in the plane of the translationally
invariant 2D homogeneous plasmonic sheet
$ (\vec{\overline{r}}_{\parallel}=\vec{r}_{\parallel}-\vec{r}_{\parallel}^{'}) $
\begin{equation}\label{2.3I1}
G_{3D}(\vec{k}_{\parallel};z,z^{\prime};\omega)=
\int{d^{2}\vec{\overline{r}}_{\parallel}}\,G_{3D}(\vec{\overline{r}}_{\parallel};z,z^{\prime};\omega)
\,{e^{-i\,\vec{k}_{\parallel}.\vec{\overline{r}}_{\parallel}}},
\end{equation}
equation (\ref{2.3I0}) becomes

\begin{equation}\label{2.3I2}
\left[\frac{\partial^{2}}{\partial{z}{^2}}+k_{\perp}^{2}\right]\,G_{3D}
(\vec{k}_{\parallel};z,z^{\prime};\omega)
= -\delta(z-z^{'})
\end{equation}
where $ k_{\perp}^{2} = q_{\omega}^{2}-k_{\parallel}^{2}$ and $ q_{\omega} = \frac{\omega}{c}\sqrt{\varepsilon_{b}^{(3D)}}$.
 This has the well-known solution [5]

\begin{equation}\label{2.4}
G_{3D}(\vec{k_\parallel};z,z^{\prime};\omega)=-\frac{e^{ik_{\perp}|z-z'|}}{2ik_{\perp}}.
\end{equation}

Employing $ G_{3D}(\vec{r},\vec{r}^{'};\omega)$, Eq.\ (\ref{2.1}) can be conveniently rewritten as an inhomogeneous integral equation as follows:
\begin{eqnarray}\label{2.5}
G_{fs}(\vec{r},\vec{r}^{'};\omega)&=&G_{3D}(\vec{r},\vec{r}^{'};\omega)
+
\frac{\omega^{2}}{c^2}\int{d^{3}\vec{r}^{''}} \int{d^{3}\vec{r}^{'''}}G_{3D}(\vec{r},\vec{r}^{''};\omega)
\alpha_{fs}^{2D}(\vec{r}^{''},\vec{r}^{'''};\omega)
G_{fs}(\vec{r}^{'''},\vec{r}^{'};\omega).
\nonumber\\
\end{eqnarray}
Introducing Eq.\  (\ref{2.2}) in Eq.\ (\ref{2.5}) and Fourier transforming the resulting equation  in the lateral plane of translational invariance ($ \vec{r}_{\parallel}-\vec{r}^{'}_{\parallel}\rightarrow \vec{k}_\parallel $), we obtain
\begin{eqnarray}\label{2.6}
G_{fs}(\vec{k}_{\parallel};z,z^{'};\omega)=
G_{3D}(\vec{k}_{\parallel};z,z^{'};\omega)
+\frac{\omega^{2}\,d\,}{c^2}
G_{3D}(\vec{k}_{\parallel};z,0;\omega)
\alpha_{fs}^{(2D)}(\vec{k}_{\parallel};\omega)
G_{fs}(\vec{k}_{\parallel};0,z^{'};\omega).
\nonumber\\
\end{eqnarray}
Solving for $ G_{fs}(\vec{k}_{\parallel};0,z^{'};\omega) $   algebraically, we obtain
\begin{widetext}
\begin{eqnarray}\label{2.7}
G_{fs}(\vec{k}_{\parallel};z,z^{'};\omega)=
G_{3D}(\vec{k}_{\parallel};z,z^{'};\omega)
+ \frac{
\omega^{2}\,d}{c^2}\,\frac{G_{3D}(\vec{k}_{\parallel};z,0;\omega)
\alpha_{fs}^{(2D)}(\vec{k}_{\parallel};\omega)
G_{3D}(\vec{k}_{\parallel};0,z^{'};\omega)}{1-\frac{
\omega^{2}}{c^2}d \alpha_{fs}^{(2D)}(\vec{k}_{\parallel};\omega)
G_{3D}(\vec{k}_{\parallel};0,0;\omega)},
\nonumber\\
\end{eqnarray}
\end{widetext}
and using Eq.\  (\ref{2.4}), this leads to the full sheet Green's function as

\begin{equation}\label{2.8}
G_{fs}(\vec{k_\parallel};z,z^{\prime};\omega)=-\frac{e^{ik_{\perp}|z-z'|}}{2ik_{\perp}}+ \frac{\gamma e^{ik_{\perp}(|z|+|z'|)}}{2ik_{\perp}(2ik_{\perp}+\gamma)},
\end{equation}
where
\begin{equation}\label{2.8I}
\gamma = \omega^2\,d\, \alpha_{fs}^{(2D)}(\vec{k}_{\parallel};\omega)/c^2.
\end{equation}

Our analysis (below) of the Green's function in the presence of an aperture will be seen to devolve upon the evaluation of $G_{fs}(\vec{r}, \vec{r}\,^{\prime\prime} ; z , z^{\prime\prime};\omega)$ at the aperture position
$ \vec{r}\,^{\prime\prime} = 0, z^{\prime\prime} = 0 $ in position representation as given by
\begin{equation}\label{2.10}
G_{fs}(\vec{r_\parallel}, 0 ; z , 0;\omega) = \frac{1}{({2\pi})^{2}}\int{d^{2}\vec{k}_{\parallel}}e^{i\vec{k}_{\parallel}.\vec{r}_{\parallel}}G_{fs}(\vec{k_\parallel};z,0;\omega)
\end{equation}
where
\begin{equation}\label{2.11}
G_{fs}(\vec{k_\parallel};z,0;\omega) = -
\frac{e^{ik_{\perp}|z|}}{2ik_{\perp}+\gamma},
\end{equation}
so that 		
\begin{equation}\label{2.12}
G_{fs}(\vec{r_\parallel}, 0 ; z , 0;\omega) = - \frac{1}{{4i\pi}}\int_{0}^{\infty}dk_{\parallel}
\ \frac{ k_{\parallel} J_{0}(k_{\parallel} r_{\parallel})e^{{i}k_{\perp}|z|}}{k_{\perp}+\frac{\gamma}{2i}}.
\end{equation}

Noting Eq.\  (\ref{2.8I}) and employing the $ 2D $ polarizability of the layer as [6]
\begin{equation}\label{2.14}
\alpha^{(2D)}_{fs}(\vec{k}_{\parallel};\omega) =\left( \frac{2\pi
i e^{2}
n_{2D}}{m^{*}\omega^{2}}\right)\sqrt{q_{\omega}^{2}-k_{\parallel}^{2}}
\end{equation}
($ n_{2D}$ is the 2D equilibrium density on the sheet, $ {m^{*}}$ is the
effective mass and $ \mu = \frac{\pi e^{2} d\, n_{2D}}{m^{*} c^{2}} $), we have

\small
\begin{eqnarray}\label{2.15}
G_{fs}(\vec{r_\parallel}, 0 ; z , 0;\omega) &=& - \frac{1}{{4i\pi}[1+{\mu}]}
  \int_{0}^{\infty}\frac{dk_{\parallel} k_{\parallel} J_{0}(k_{\parallel} r_{\parallel})e^{{i}|z|\sqrt{q_{\omega}^{2}-k_{\parallel}^{2}}}}{\sqrt{q_{\omega}^{2}-k_{\parallel}^{2}}}.
\nonumber\\
\end{eqnarray}

 The $ k_{\parallel}$ - integral of Eq.\ (\ref{2.15}) is readly evaluated as[7]
\begin{equation}\label{2.21}
 G_{fs}(\vec{r_\parallel}, 0 ; z , 0;\omega) =  \frac{1}{{4\pi}[1+{\mu}]}
\frac{e^{{i}\,q_{\omega}\sqrt{r_{\parallel}^{2}+|z|^{2}}}}{\sqrt{r_{\parallel}^{2}+|z|^{2}}}.
\end{equation}

\newpage
\section{Green's function solution for a Perforated 2D plasmonic layer with a nano-hole embedded in a 3D bulk host medium}
\label{sec3}
\subsection{Integral Equation: Scalar Helmholtz Green's Function for a Perforated Plasmonic Layer}
\begin{figure}[h]
\centering
\includegraphics[width=17cm,height=7cm]{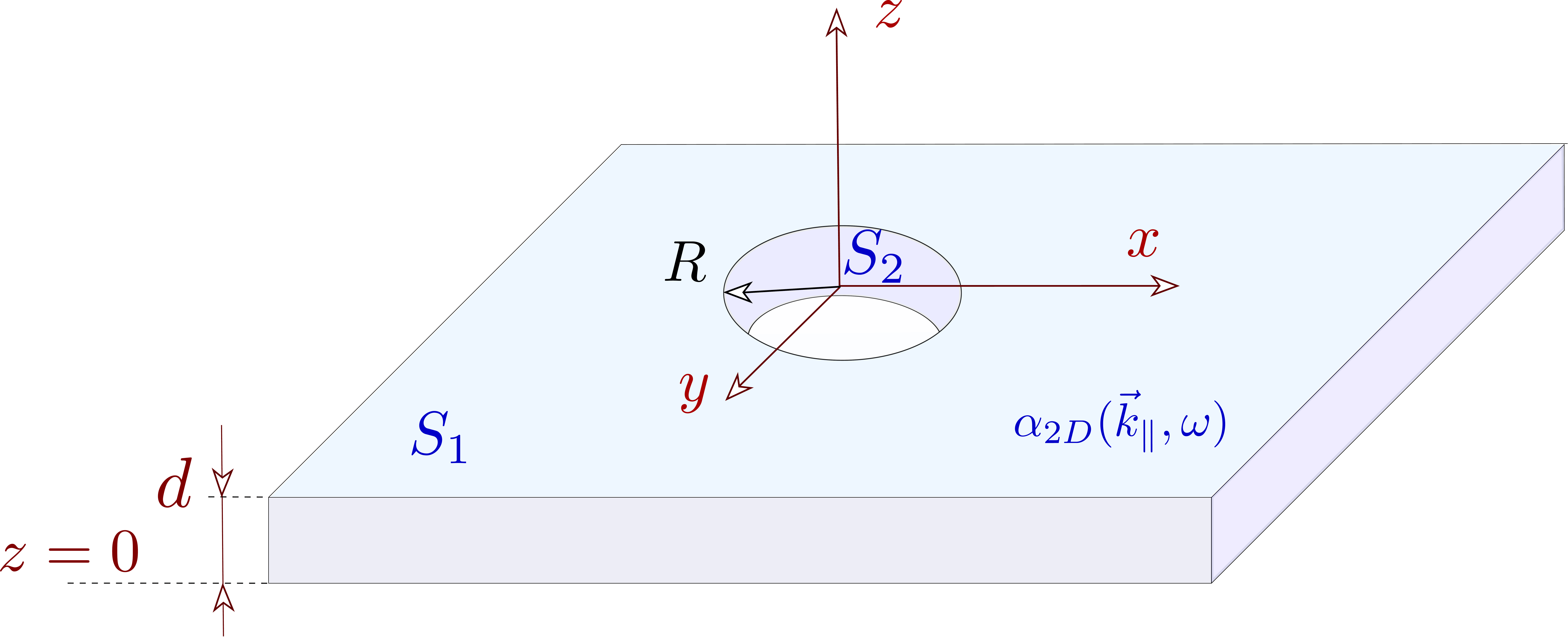}
\caption{(Color online)\  Schematic representation of a perforated 2D plasmonic layer (thickness $ d $, embedded
at $ z=0 $ in a three dimensional bulk meduim) with a nano-hole of radius
$ R $ at the origin of the $(x-y)$-plane.}
\label{FIG311R}
\qquad
\end{figure}

We consider a 2D plasmonic layer $ S_{1} $  which is perforated by a nano-scale aperture $ S_{2} $ of radius $ R $, as depicted in Fig.\ref{FIG311R}, lying in the  $(x-y)$-plane.  The presence of the nano-hole in the layer is represented by subtracting the part of the polarizability associated with the hole from the polarizability of the full layer, (Fig.\ref{FIG311R})[1,2,3,4]
\begin{equation}\label{3.1}
\alpha^{(2D)}(\vec{r},\vec{r}^{'};\omega)=\alpha^{(2D)}_{fs}(\vec{r},\vec{r}^{'};\omega)
-\alpha^{(2D)}_{h}(\vec{r},\vec{r}^{'};\omega)
\end{equation}
where $ \alpha^{(2D)}_{h}(\vec{r},\vec{r}^{'};\omega) $ is the part of the layer polarizability
removed by the nano-hole.  The resulting Green's function for the perforated plasmonic layer with the hole
satisfies the integral equation given by
\begin{eqnarray}\label{3.2}
G(\vec{r},\vec{r}^{'};\omega)&=&G_{fs}(\vec{r},\vec{r}^{'};\omega)
-
\frac{\omega^{2}}{c^2}\int{d^{3}\vec{r}^{''}} \int{d^{3}\vec{r}^{'''}}G_{fs}(\vec{r},\vec{r}^{''};\omega)
 \times
\alpha_{h}^{2D}(\vec{r}^{''},\vec{r}^{'''};\omega)G(\vec{r}^{'''},\vec{r}^{'};\omega).
\nonumber\\
\end{eqnarray}
Here, the polarizability of the nano-hole is defined as [8,9]
\begin{eqnarray}\label{3.3}
\alpha^{(2D)}_{h}(\vec{r},\vec{r}^{'};\omega)&=&\alpha^{(2D)}_{fs}(x,y,x^{'},y^{'};\omega)
\eta_{+}(R-|x|)\eta_{+}(R-|y|)
\eta_{+}(R-|x^{'}|)\eta_{+}(R-|y|^{'})
d \, \delta(z) \delta(z'),
\nonumber\\
\end{eqnarray}
where $ \eta_{+}(R-|x|)$ is the Heaviside unit step function representing a cut-off imposed to confine the integration range on the 2D sheet to the nano-hole dimensions; and the Dirac delta function $ \delta(z)$ is needed to localize the polarizability onto the plane of the 2D plasmonic layer.
A simple approximation of Eq.\  (\ref{3.3}) for very small radius leads to (A represents the area of the aperture)
\begin{eqnarray}\label{3.4}
\alpha^{(2D)}_{h}(\vec{r},\vec{r}^{'};\omega)&\approx & A^2 \alpha^{(2D)}_{fs}(\vec{r_\parallel},\vec{r_\parallel}^{'};\omega)
\delta^{2D}(\vec{r_\parallel})\delta^{2D}(\vec{r_\parallel}^{'})
d \delta(z) \delta(z'),
\end{eqnarray}
and employing it  in Eq.\ (\ref{3.2}) to execute all positional integrations,
we obtain
\begin{eqnarray}\label{3.7}
G(\vec{r_{\parallel}},\vec{r_{\parallel}}^{'};z,z^{'};\omega) &\approx & G_{fs}(\vec{r_{\parallel}},\vec{r_{\parallel}}^{'};z,z^{'};\omega)
- \beta\,G_{fs}(\vec{r_{\parallel}},0;z,0;\omega)
G(0,\vec{r_\parallel}^{'};0,z^{'};\omega),
\nonumber\\
\end{eqnarray}
where $ \beta = \gamma A^{2}$ with $ \gamma =\frac{\omega^{2} d}{c^{2}} \alpha_{fs}^{2D}([\vec{r}_{\parallel}=0]-[\vec{r}_{\parallel}^{'}=0];\omega)$.
To solve Eq.\  (\ref{3.7}), we set $ \vec{r_{\parallel}}=0 $ and $ z =0 $, and
determine $ G(0,\vec{r_\parallel}^{'};0,z^{'};\omega)$ as
\begin{equation}\label{3.8}
G(0,\vec{r_{\parallel}}^{'};0,z^{'};\omega) =  \frac{G_{fs}(0,\vec{r_{\parallel}}^{'};0,z^{'};\omega)}{1+\beta\,G_{fs}(0,0;0,0;\omega)}\  .
\end{equation}
Substituting Eq.\  (\ref{3.8}) into Eq.\  (\ref{3.7}) yields the algebraic closed form analytic solution:
\begin{eqnarray}\label{3.9}
G(\vec{r}_{\parallel},\vec{r}_{\parallel}^{'};z,z^{'};\omega) \approx G_{fs}(\vec{r}_{\parallel},\vec{r}_{\parallel}^{'};z,z^{'};\omega)
-\,\,\frac{\beta G_{fs}(\vec{r}_{\parallel},0;z,0;\omega)
G_{fs}(0,\vec{r}_{\parallel}^{'};0,z^{'};\omega)}{1+\beta
G_{fs}([\vec{r}_{\parallel}=0]-[\vec{r}_{\parallel}^{'}=0];z=0,z^{'}=0;\omega)}.
\nonumber\\
\end{eqnarray}
\normalsize
Noting that a transmitted scalar wave is controlled by $ G(\vec{r}_{\parallel},0;z,0;\omega)$ , we have analyzed this quantity setting $ \vec{r}_{\parallel}^{'} = 0 $ and $ z^{'}=0 $ in Eq.\  (\ref{3.9}), leading to
\begin{equation}\label{3.10}
G(\vec{r}_{\parallel},0;z,0;\omega)\approx \frac{ G_{fs}(\vec{r}_{\parallel},0;z,0;\omega)
}{1+\beta G_{fs}([\vec{r}_{\parallel}=0]-[\vec{r}_{\parallel}^{'}=0];z=0,z^{'}=0;\omega)}.
\end{equation}
Eq.\  (\ref{3.10}) is an approximate analytical Green's function solution to Eq.\  (\ref{3.2}), in which\\ $
G_{fs}([\vec{r}_{\parallel}=0]-[\vec{r}_{\parallel}^{'}=0];z=0,z^{'}=0;\omega)$
is found to involve a divergent integral when all its arguments vanish.  This divergence may be seen in $ \alpha^{(2D)}_{fs} $ and $ G_{fs} $ setting $ \vec{r}_{\parallel}=0 $,  $ \vec{r}_{\parallel}^{'}=0 $,
 $ z = 0 $ and $ z^{'}=0 $ as follows:
\begin{equation}\label{3.11}
\alpha^{(2D)}_{fs}(0,0;\omega) =\frac{1}{2\pi}\left( \frac{2\pi i e^{2} n_{2D}}{m^{*}\omega^{2}}\right)\int_{0}^{\infty}dk_{\parallel} k_{\parallel}\sqrt{q_{\omega}^{2}-k_{\parallel}^{2}},
\end{equation}
and Eq.\  (\ref{2.15}) is given by
\begin{equation}\label{3.12}
G_{fs}(0, 0 ; 0 , 0;\omega) = \frac{i}{{4\pi}[1+{\mu}]}\int_{0}^{\infty}\frac{dk_{\parallel} k_{\parallel}}{\sqrt{q_{\omega}^{2}-k_{\parallel}^{2}}}.
\end{equation}
The divergence of $ \alpha^{(2D)}_{fs} $ and $ G_{fs} $ is an artifact of limiting the radius of the aperture to be vanishingly small (zero) in the kernel of the integral equation,
Eq.\  (\ref{3.2}) -  Eq.\  (\ref{3.4}).  A more realistic consideration involves a cut off at a small but finite radius $ R $, alternatively an upper limit on the wavenumber integration, $ k_{\parallel}\sim \frac{1}{R}$, yielding the convergent integrals
\begin{equation}\label{3.19}
\alpha^{(2D)}_{fs}(0,0;\omega) =\frac{q_{\omega}^{3}}{2\pi}\left( \frac{2\pi i e^{2} n_{2D}}{m^{*}\omega^{2}}\right)\\
\left\{\int_{0}^{1}dy y\sqrt{1-y^{2}}+i\,
\int_{1}^{\frac{1}{q_{\omega}\,R}}dy y\sqrt{y^{2}-1}
\right\}
\end{equation}
and
\small
\begin{equation}\label{3.25}
G_{fs}([\vec{r}_{\parallel}=0]-[\vec{r}_{\parallel}^{'}=0];z=0,z^{'}=0;\omega)
\approx
\frac{\,i\,q_{\omega}}{{4\pi}[1+{\mu}]}\\
\left\{\int_{0}^{1}dy\ \frac{y}{\sqrt{1-y^{2}}}
-\,i\, \int_{1}^{\frac{1}{q_{\omega} R}}dy\  \frac{y}{\sqrt{y^{2}-1}}\right\},
\end{equation}
\normalsize
where we have introduced the dimensionless notation $ y=k_{\parallel} / q_{\omega} $,  so that
\begin{equation}\label{3.18}
k_{\perp}=
\left\{
  \begin{array}{ll}
    q_{\omega}\,{\sqrt{1-y^{2}}} & \hbox{  for $ 0\leq y < 1 $ ;} \\
    i\, q_{\omega}\,{\sqrt{y^{2}-1}} & \hbox{ for $ y > 1 $ .}
  \end{array}
\right.
\end{equation}
and the following results are obtained
\begin{equation}\label{3.21}
\alpha^{(2D)}_{fs}([\vec{r}_{\parallel}=0],[\vec{r}^{'}_{\parallel}=0];\omega)\approx
\frac{1}{6\pi
R^{3}} \left( \frac{2\pi i e^{2}
n_{2D}}{m^{*}\omega^{2}}\right)\\
\left[\left(q_{\omega}R
\right)^{3}+i\,\left[1-(q_{\omega}R)^{2}\right]^{3/2}\right]
\end{equation}
and
\begin{equation}\label{3.28}
G_{fs}([\vec{r}_{\parallel}=0]-[\vec{r}_{\parallel}^{'}=0];z=0,z^{'}=0;\omega)\approx \frac{1}{4\pi(1+\mu)R}\\
\left[\,i\,q_{\omega}R \,+
\,\sqrt{1-(q_{\omega}R)^{2}}\right],
\end{equation}
for $ q_{\omega}R < 1 $ .
Furthermore, substituting Eq.\  (\ref{3.21}) into the expression for $ \gamma $, we have
\begin{equation}\label{3.23}
\gamma=\frac{\mu}{3\pi R^{3}} \left[\,i\,\left(\,q_{\omega}R
\right)^{3}\,-\,\left[1-(q_{\omega}R)^{2}\right]^{3/2}\right]
\end{equation}
with $ \mu = \pi e^{2} d\, n_{2D}/(m^{\ast} c^{2})$ .

\section{ Numerical Results}
\label{sec4}
\begin{figure*}[h]
\centering
\includegraphics[width=8.0cm,height=6cm]{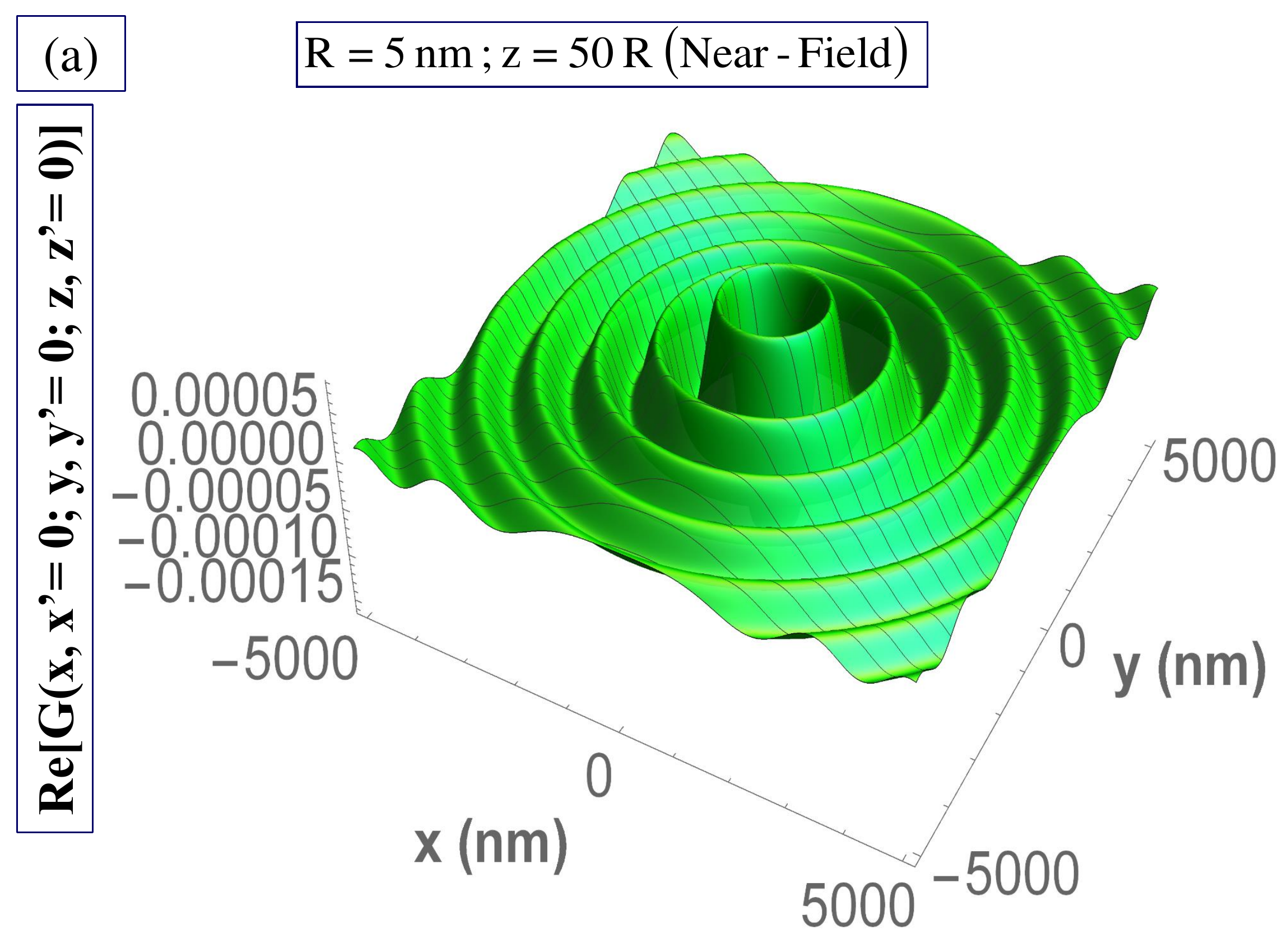}
\includegraphics[width=8.0cm,height=6cm]{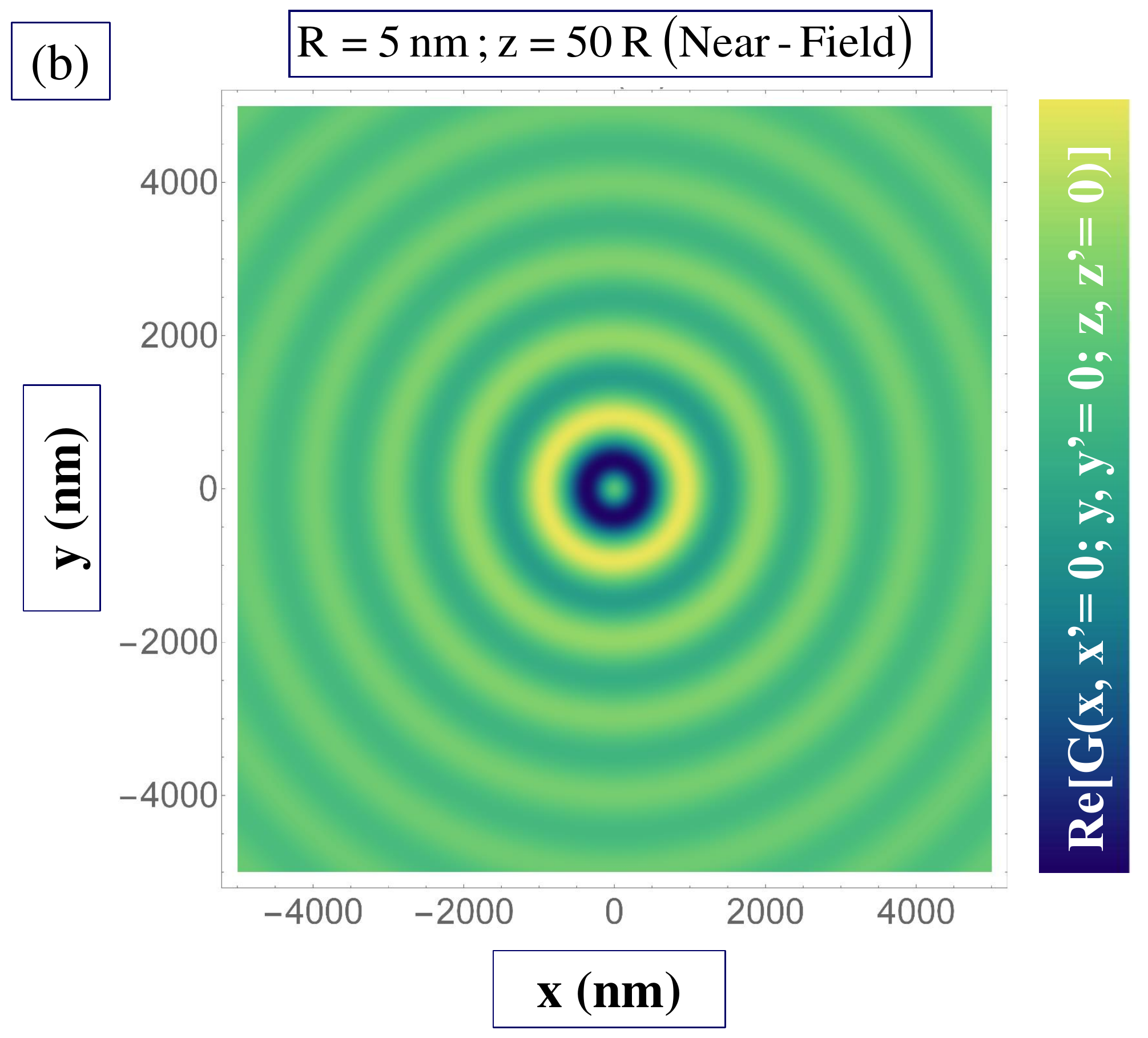}\\
\includegraphics[width=8.0cm,height=6cm]{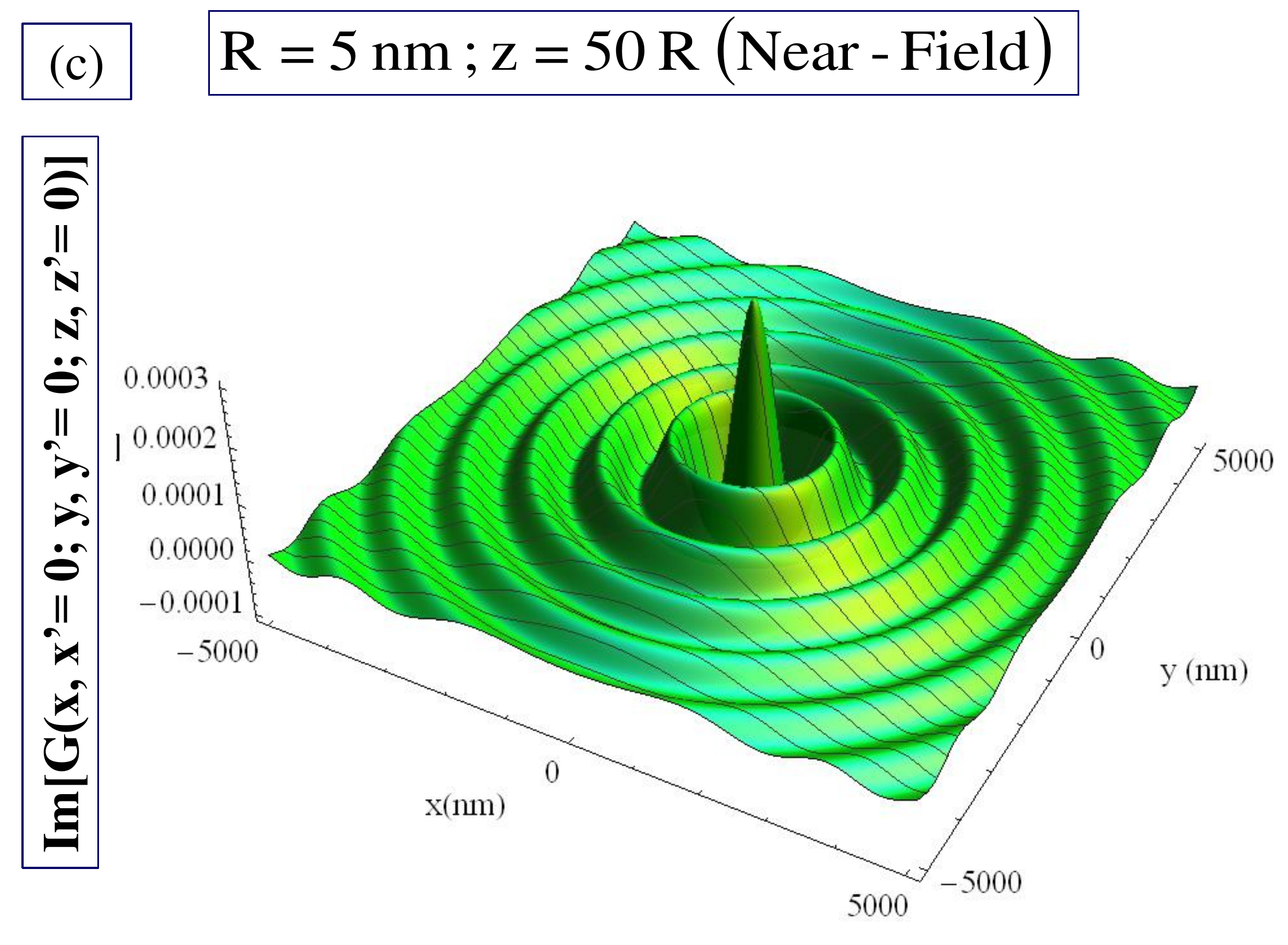}
\includegraphics[width=8.0cm,height=6cm]{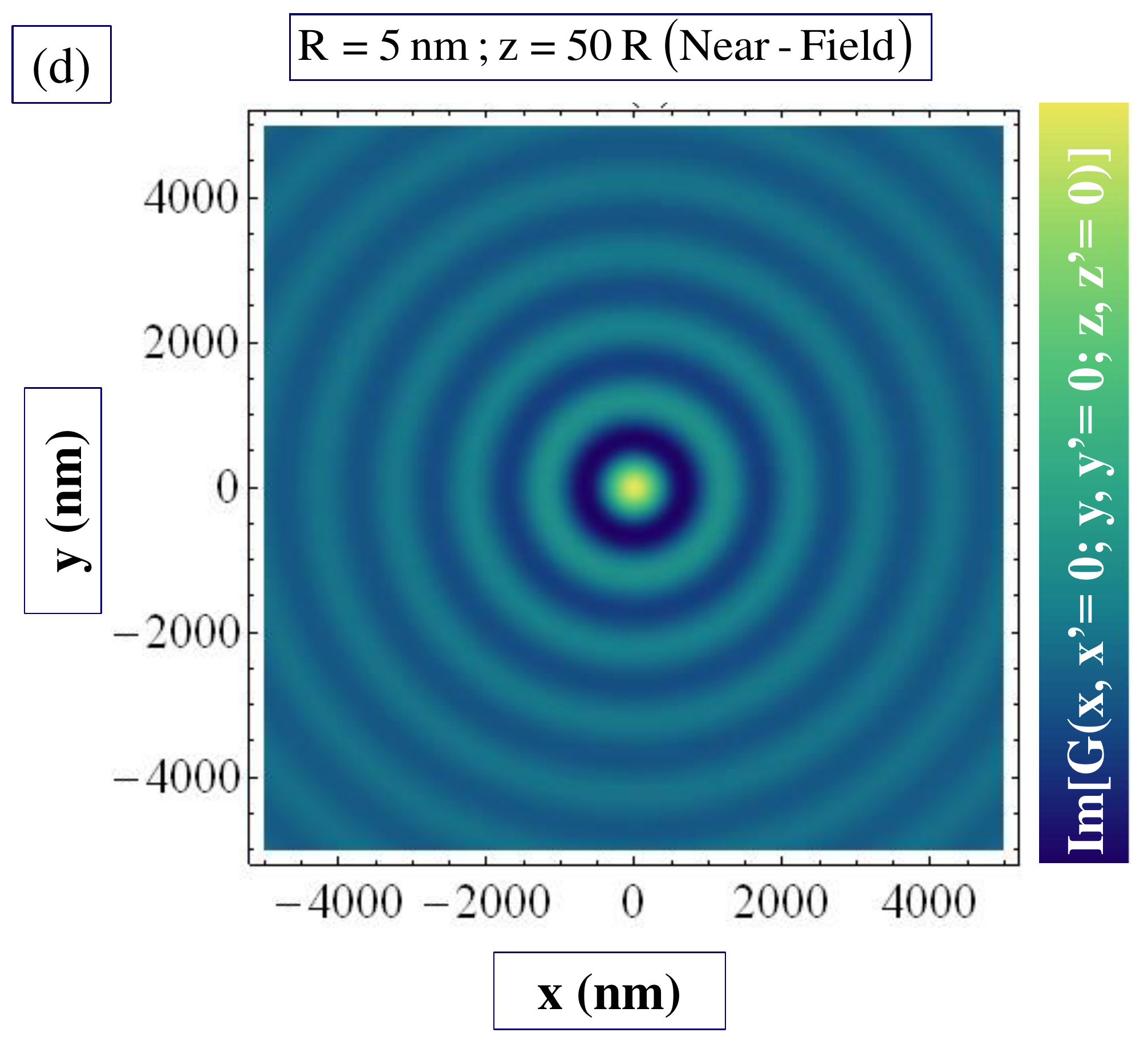}
\caption{(Color online)\ Figures  $ (a)$ and $( b)$ exhibit Re$[G(\vec{r}_{\parallel},0;z,0;\omega)]$ and
 $ (c)$ and $ (d)$  present Im$[G(\vec{r}_{\parallel},0;z,0;\omega)] $ (in 3D (a, c) and density (b, d) plots) for a perforated 2D plasmonic layer of GaAs in the presence of a nano-hole of radius $ R=5\,nm $ at $ z = 50\,R $ (Near-Field) for $ \varepsilon_{b}=1 $, $ n_{_{2D}}=4\times10^{15}cm^{-2} $, $ d=10\,nm $ and $ m^{\ast}=0.065 m_{0} $ where $ m_{0} $ is the free-electron mass.}
\label{3DNFR5GRe1}
\qquad
\end{figure*}

\begin{figure*}[h]
\centering
\includegraphics[width=8cm,height=6cm]{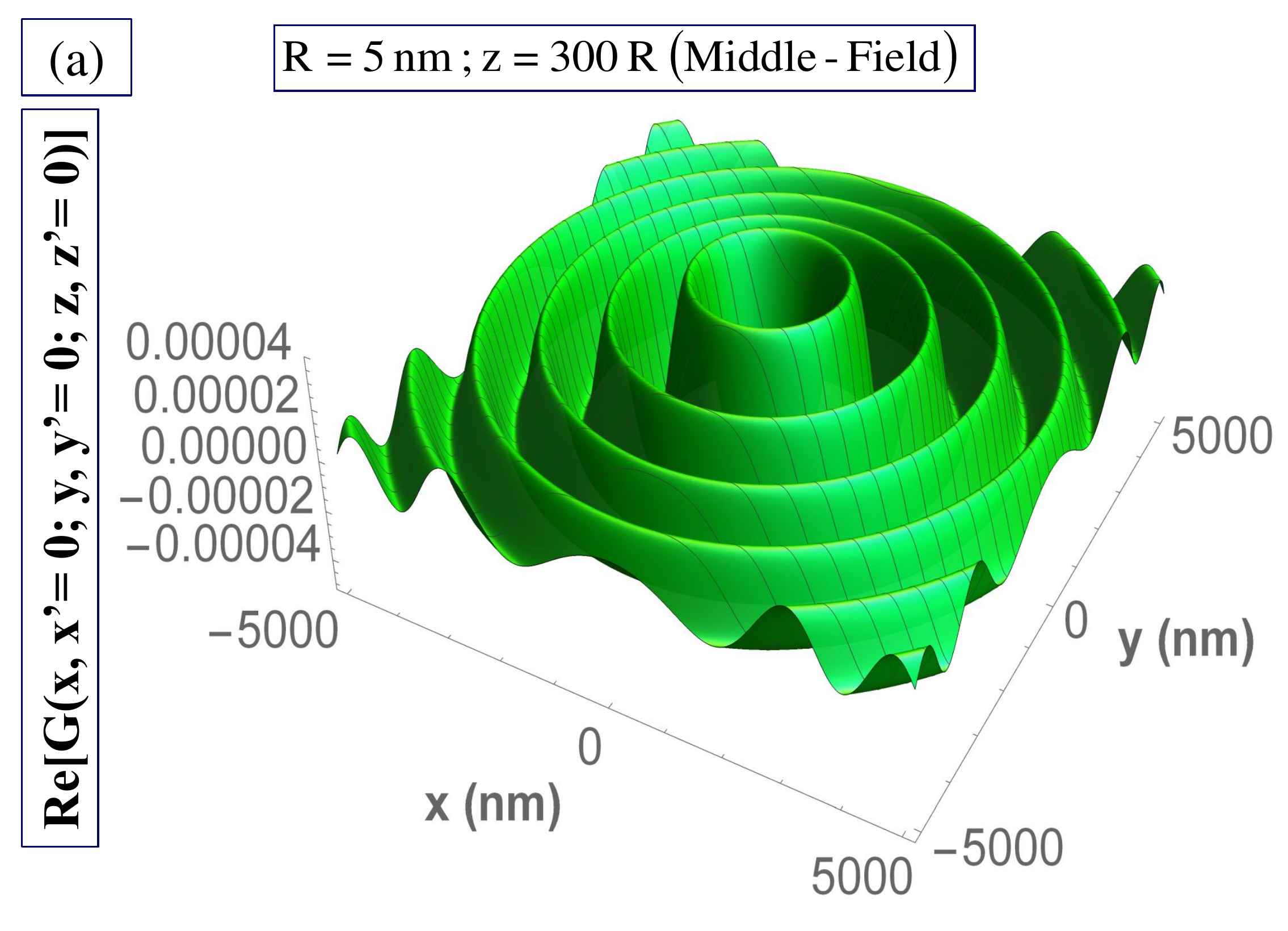}
\includegraphics[width=8cm,height=6cm]{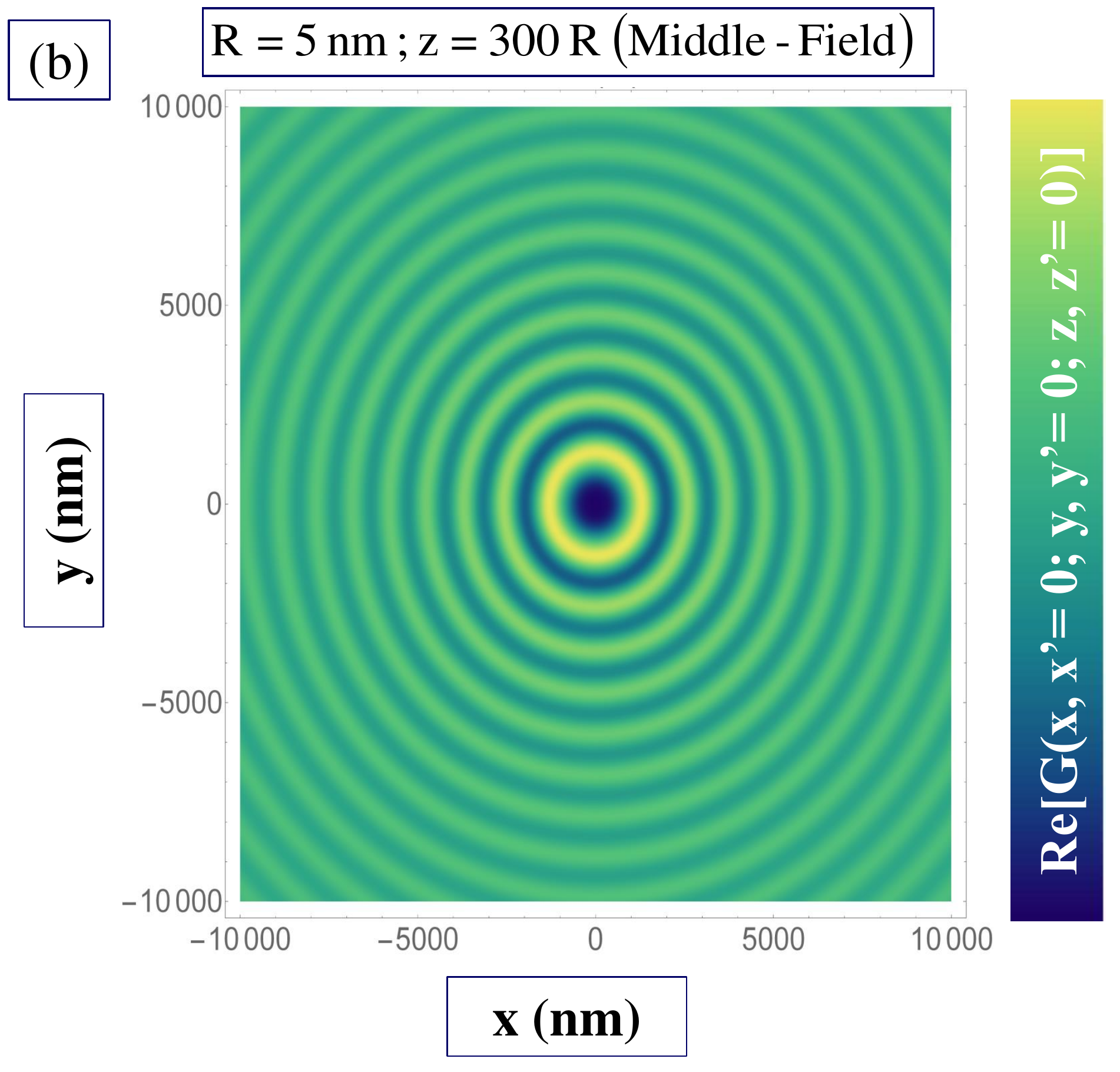}\\
\includegraphics[width=8cm,height=6cm]{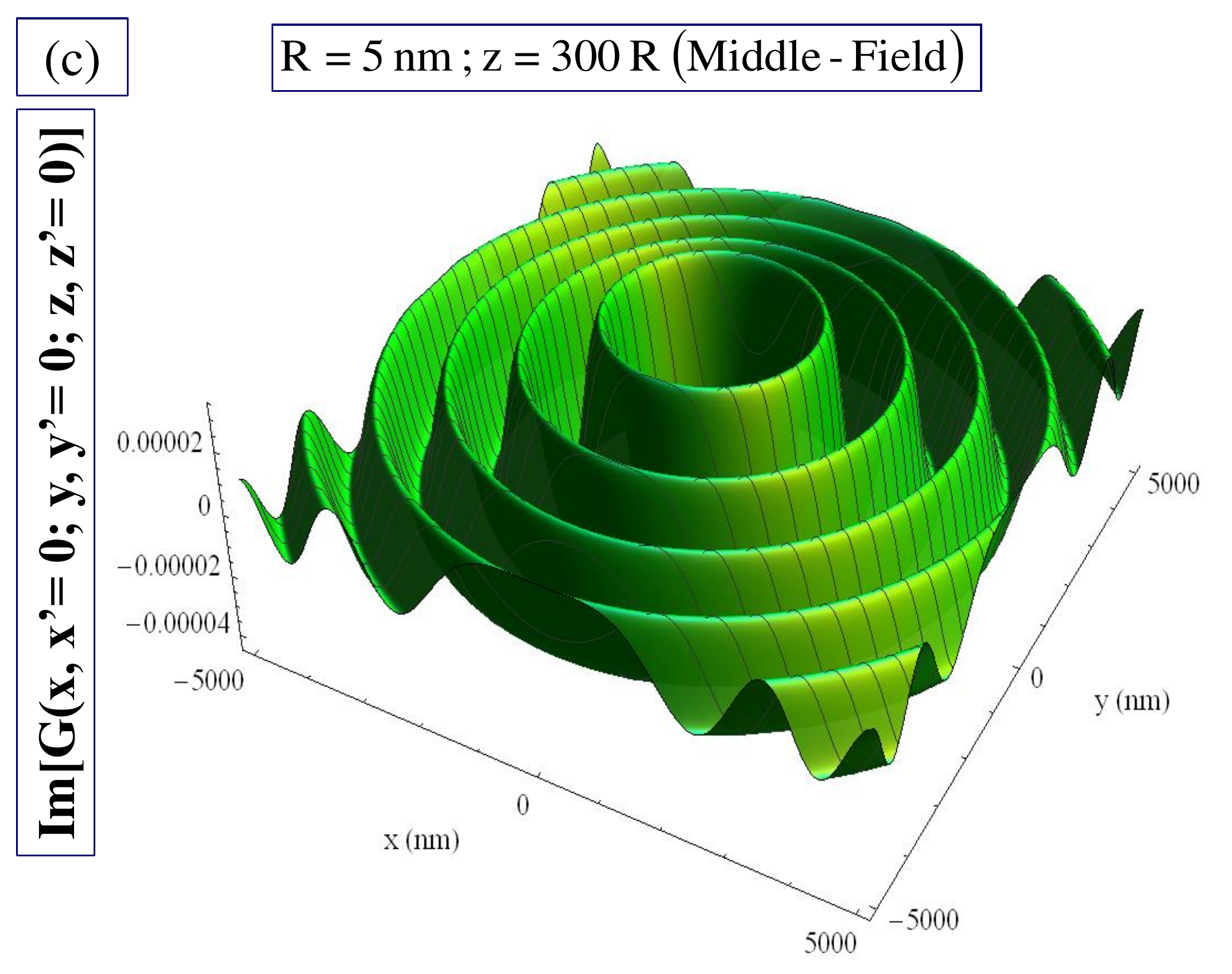}
\includegraphics[width=8cm,height=6cm]{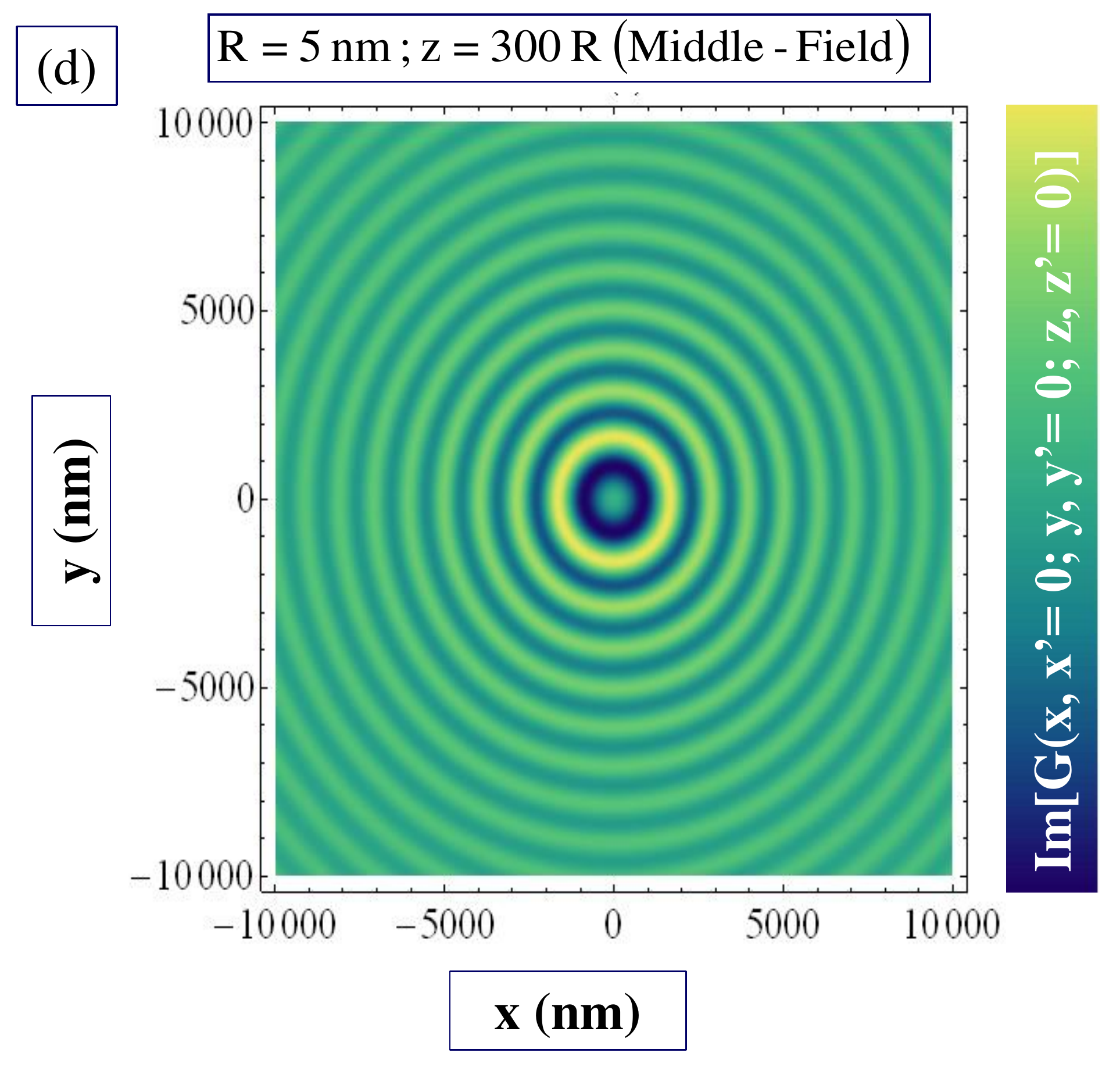}
\caption{(Color online)\ Figures  $ (a)$ and $( b)$ exhibit Re$[G(\vec{r}_{\parallel},0;z,0;\omega)]$ and
 $ (c)$ and $ (d)$  present Im$[G(\vec{r}_{\parallel},0;z,0;\omega)] $ (in 3D (a, c) and density (b, d) plots) for a perforated 2D plasmonic layer of GaAs in the presence of a nano-hole of radius $ R=5\,nm $ at $ z = 300\,R $ (Middle-Field) for $ \varepsilon_{b}=1 $, $ n_{_{2D}}=4\times10^{15}cm^{-2} $, $ d=10\,nm $ and $ m^{\ast}=0.065 m_{0} $ where $ m_{0} $ is the free-electron mass.}
\label{3DMFR5GRe1}
\qquad
\end{figure*}

\begin{figure*}[tbh]
\centering
\includegraphics[width=8cm,height=6cm]{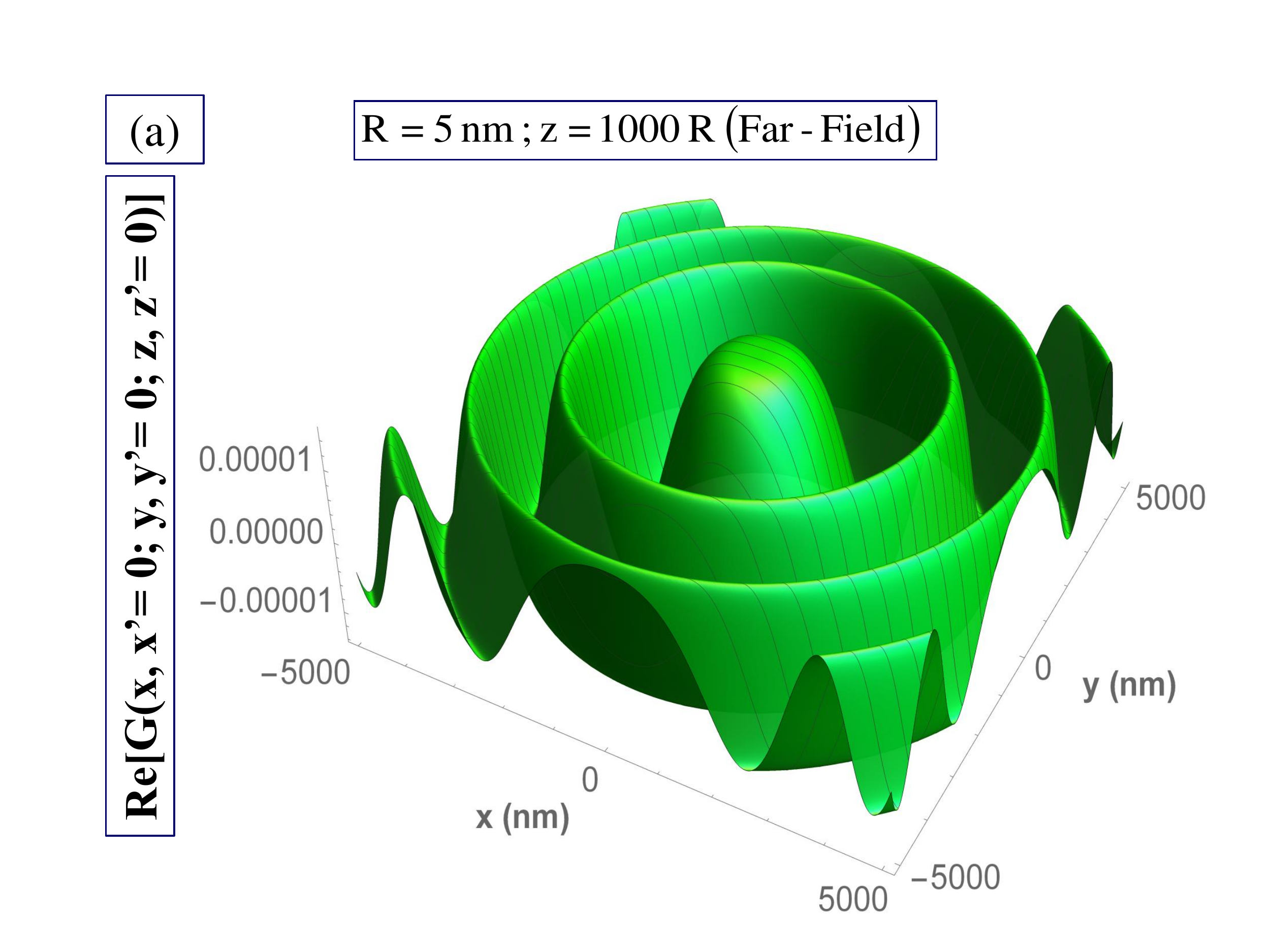}
\includegraphics[width=8cm,height=6cm]{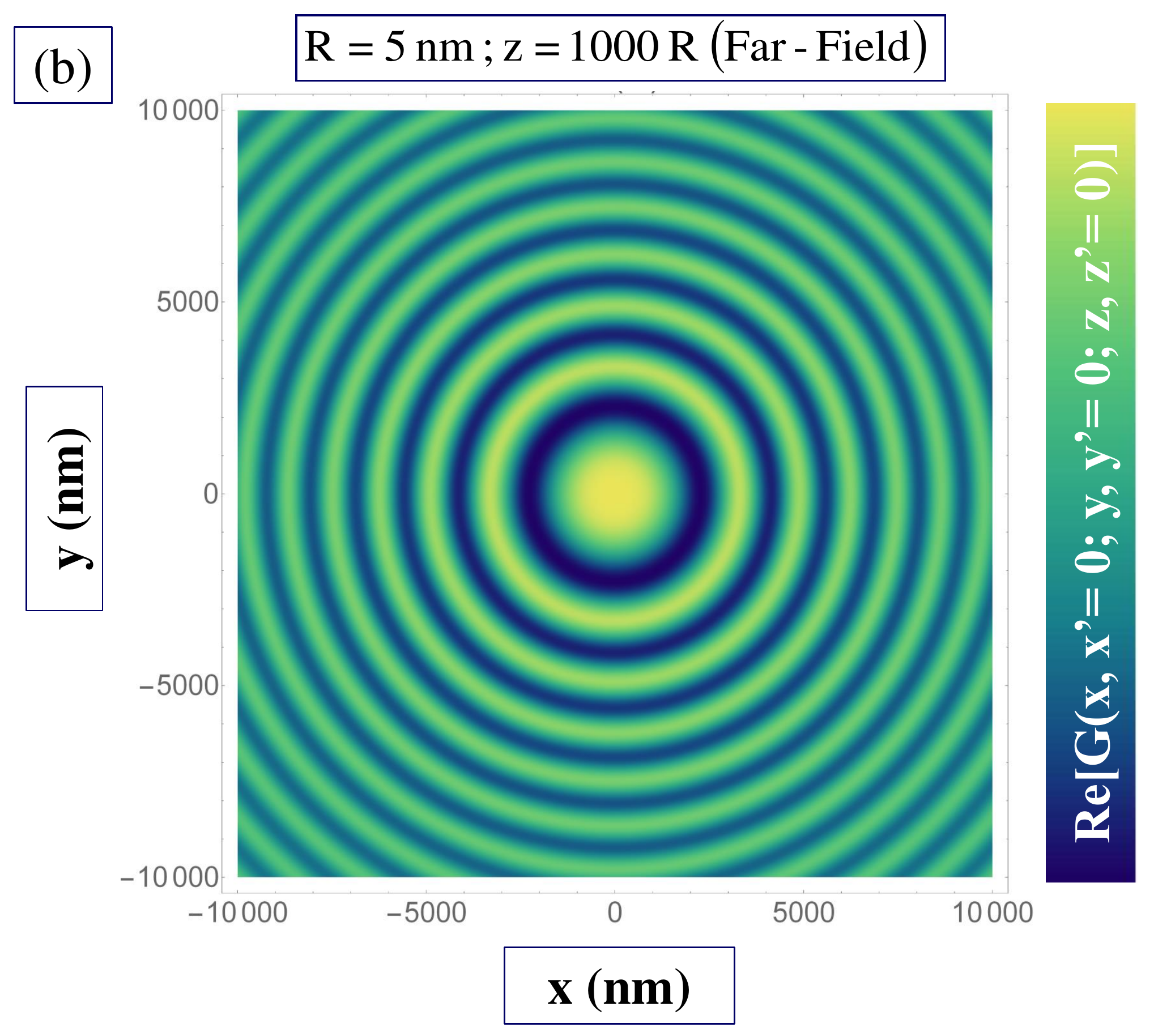}\\
\includegraphics[width=8cm,height=6cm]{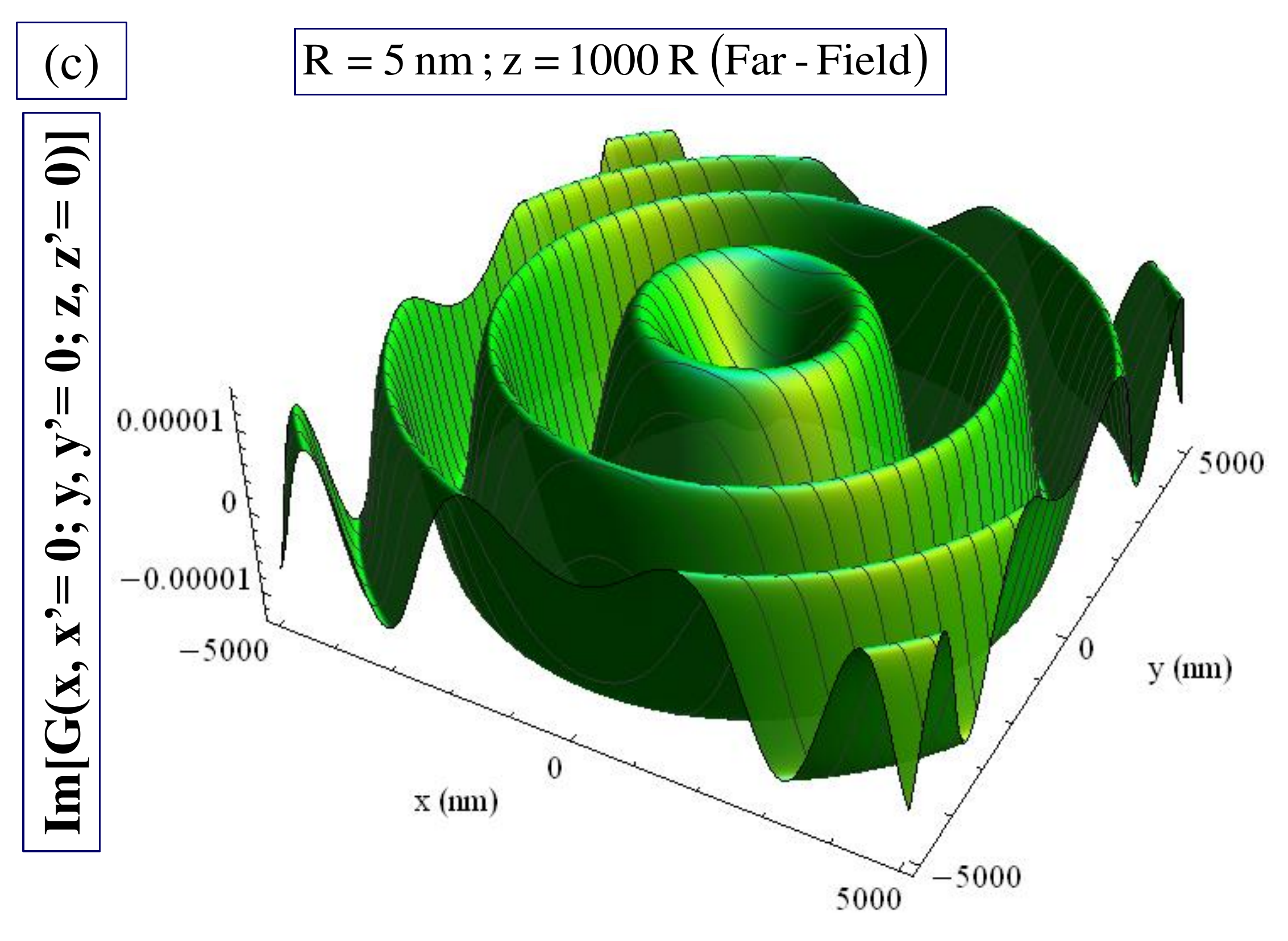}
\includegraphics[width=8cm,height=6cm]{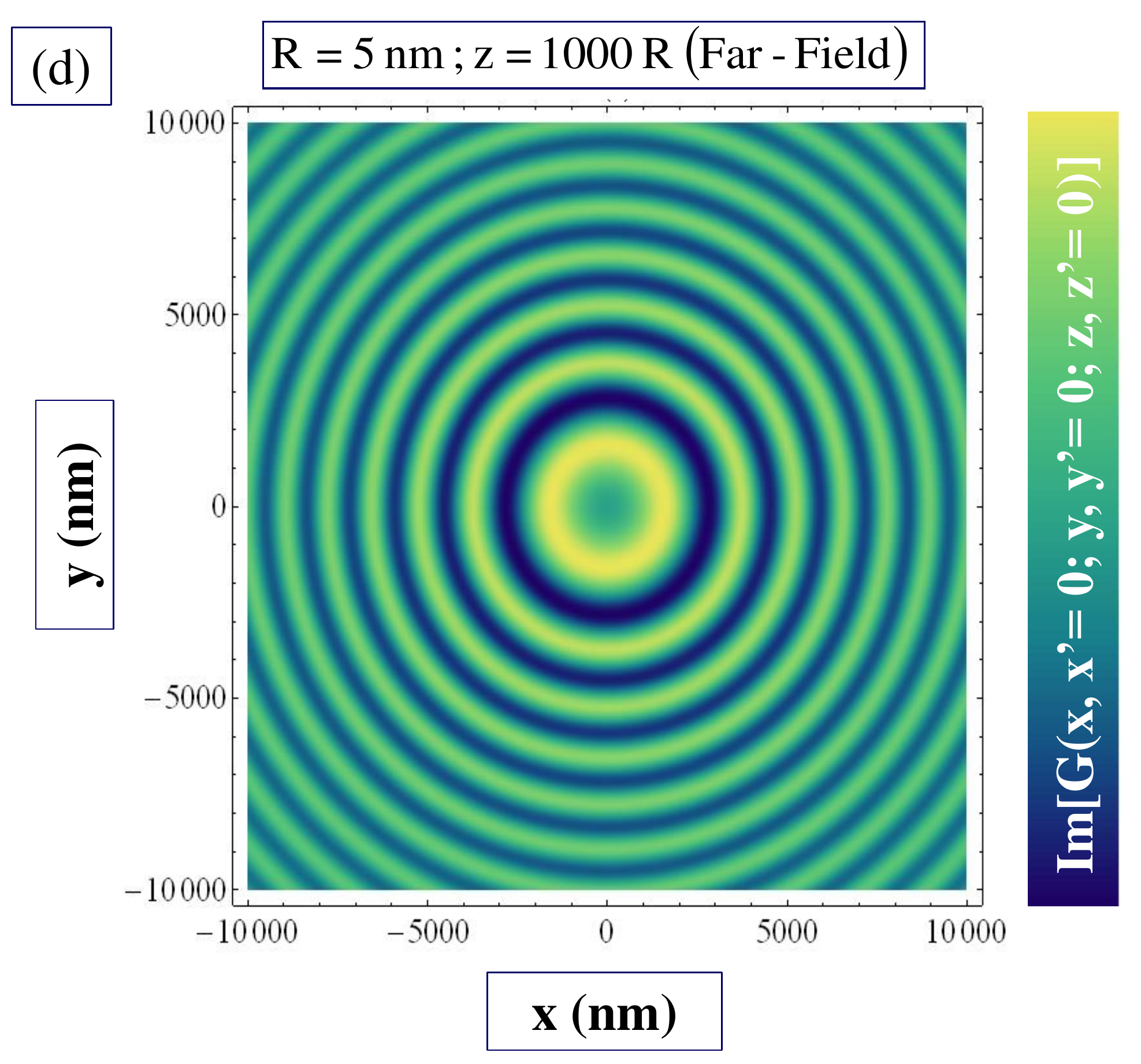}
\caption{(Color online)\ Figures  $ (a)$ and $( b)$ exhibit Re$[G(\vec{r}_{\parallel},0;z,0;\omega)]$ and
 $ (c)$ and $ (d)$  present Im$[G(\vec{r}_{\parallel},0;z,0;\omega)] $ (in 3D (a, c) and density (b, d) plots) for a perforated 2D plasmonic layer of GaAs in the presence of a nano-hole of radius $ R=5\,nm $ at $ z = 1000\,R $ (Far-Field) for $ \varepsilon_{b}=1 $, $ n_{_{2D}}=4\times10^{15}cm^{-2} $, $ d=10\,nm $ and $ m^{\ast}=0.065 m_{0} $ where $ m_{0} $ is the free-electron mass.}
\label{GR100Z2R}
\qquad
\end{figure*}

\numberwithin{figure}{section}
Our numerical results for the real and imaginary parts of the Green's function in Eq.\  (\ref{3.10}),
Re$ [G(\vec{r}_{\parallel},0;z,0;\omega)] $ and Im$ [G(\vec{r}_{\parallel},0;z,0;\omega)] $, respectively, for  frequency  $f = 300 $ THz are presented in  Figs:\ \ref{3DNFR5GRe1}, \ref{3DMFR5GRe1} and \ref{GR100Z2R}  as functions of $ x $ and $ y $ for several values of distance $ z $ away from the layer screen: We chose $ z = 50\,R $  (near-field),  $ z = 300\,R $ (middle-field) and  $ z = 1000\,R $  (far-field).  These figures reveal the structure of the Green's function  for the perforated layer in terms of near-field ( $ z = 50\,R $ ),
middle-field ($ z = 300\,R $) and far-field ($ z = 1000\,R $) radiation zones for $ R=5\,nm $.
\newpage
\newpage
Further detail concerning Re$ [G(\vec{r}_{\parallel},0;z,0;\omega)]$ and  Im$ [G(\vec{r}_{\parallel},0;z,0;\omega)] $ is provided in the figures below for the three $z$-radiation zones described by $ z = 50\,R $ (near-field), $ z = 300\,R $ (middle-field) and $ z = 1000\,R $ (far-field):
\begin{figure*}[htbp]
\centering
\includegraphics[width=8cm,height=7cm]{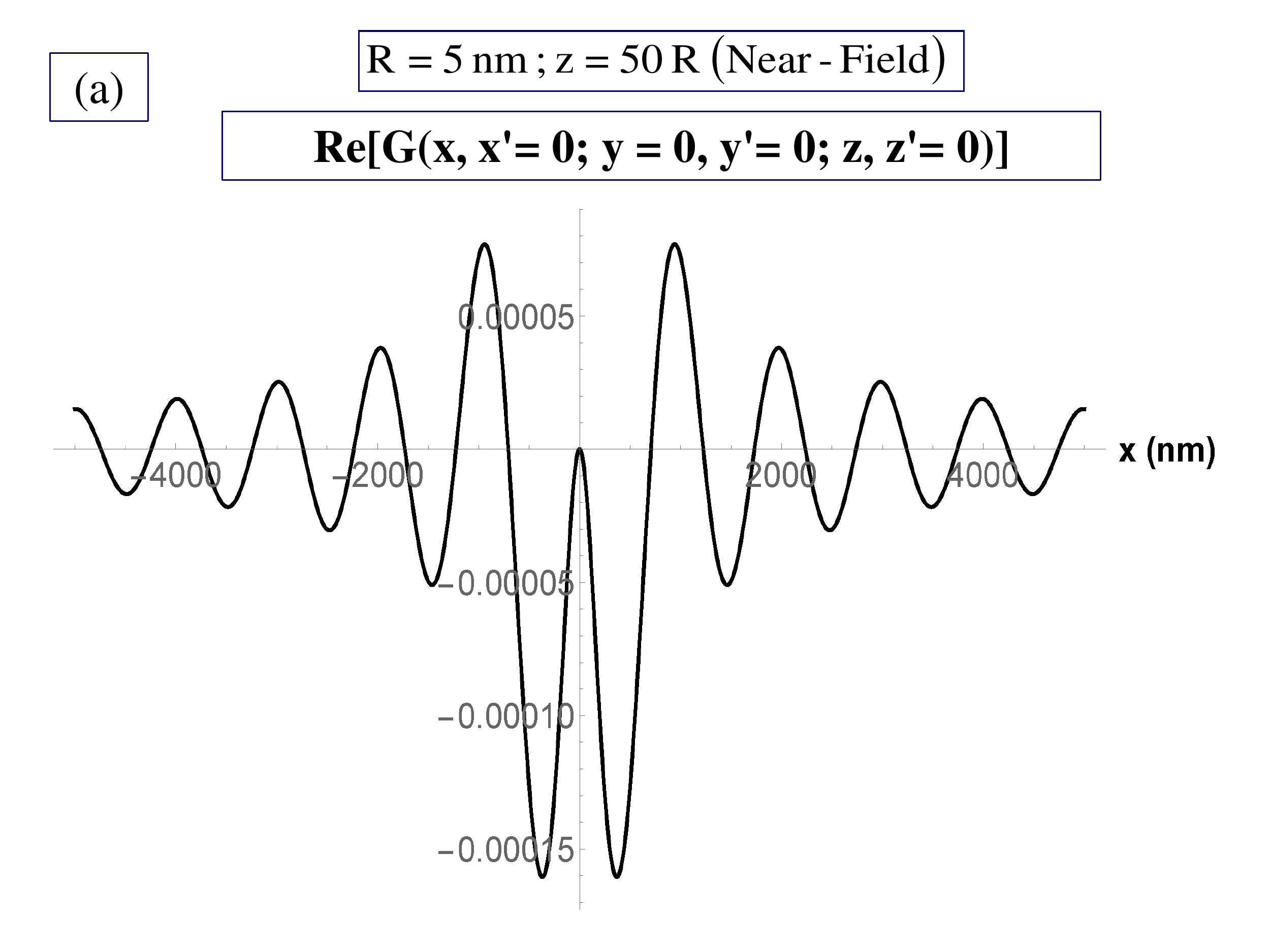}
\includegraphics[width=8cm,height=7cm]{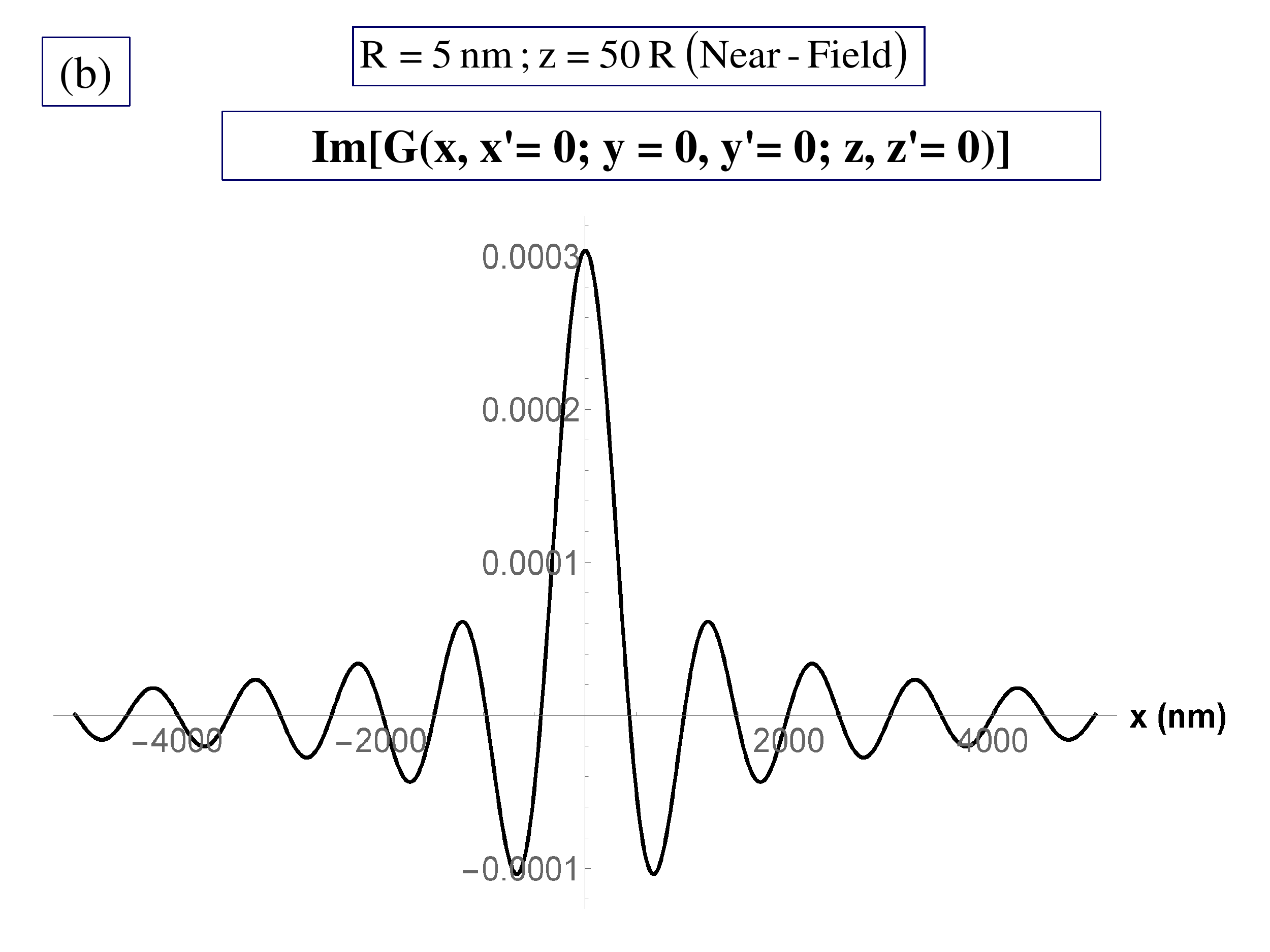}
\caption{$(a)$  exhibits Re$[G(x, x^{\prime}=0; y=0, y^{\prime}=0;z=50\,R, z^{\prime}=0;\omega)]$ and $(b)$ presents \\
Im $[G(x, x^{\prime}=0; y=0, y^{\prime}=0;z=50\,R, z^{\prime}=0;\omega)]$ as a function of $ x $ for a perforated 2D plasmonic layer of GaAs in the presence of a nano-hole of radius $ R=5\,nm $ at $ z = 50\,R $ (Near-Field) for $ \varepsilon_{b}=1 $, $ n_{_{2D}}=4\times10^{15}cm^{-2} $, $ d=10\,nm $ and $ m^{\ast}=0.065 m_{0} $ where $ m_{0} $ is the free-electron mass.}
\label{NFR5GF}
\qquad
\end{figure*}

\begin{figure*}[htbp]
\centering
\includegraphics[width=8cm,height=7cm]{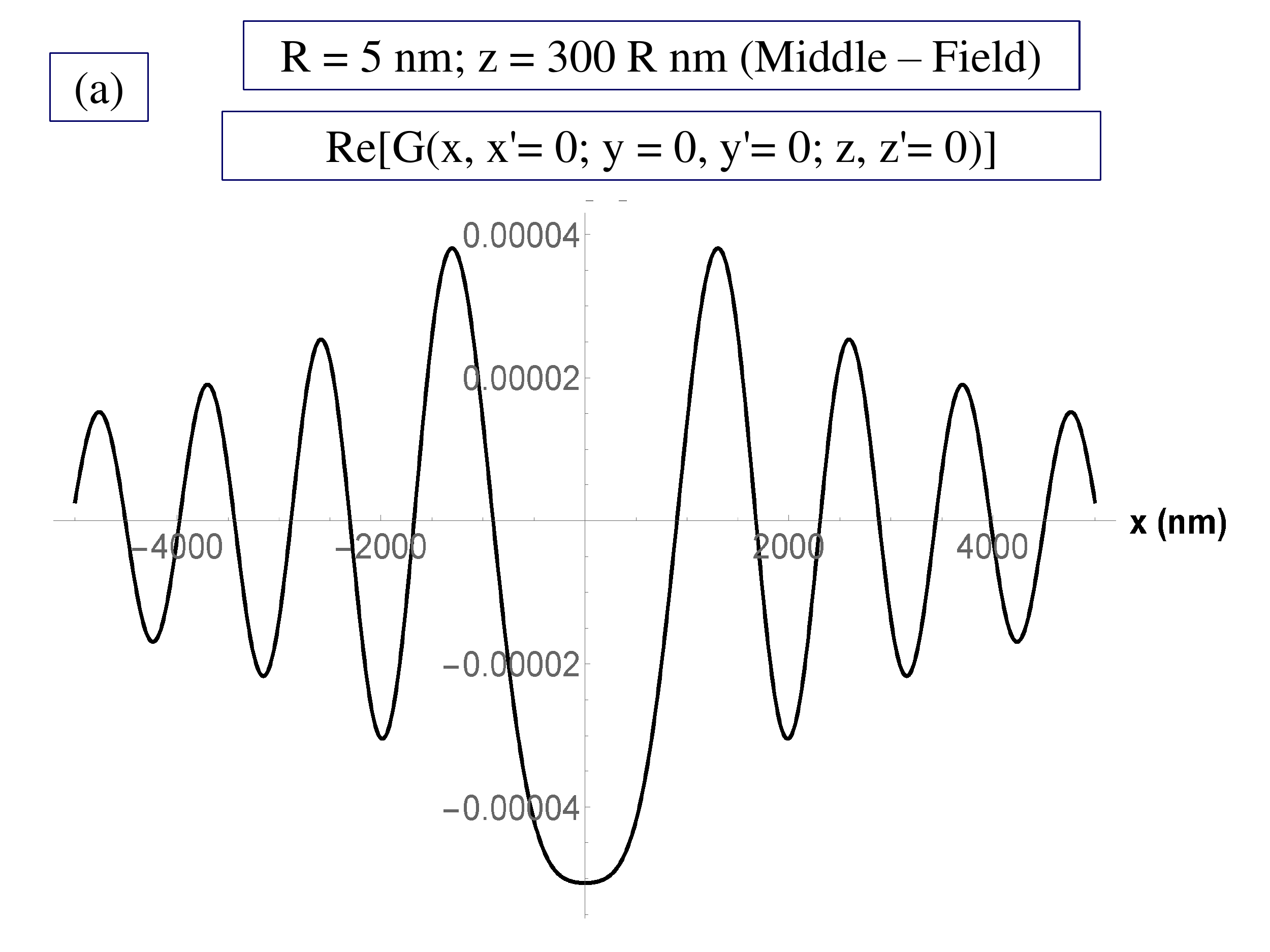}
\includegraphics[width=8cm,height=7cm]{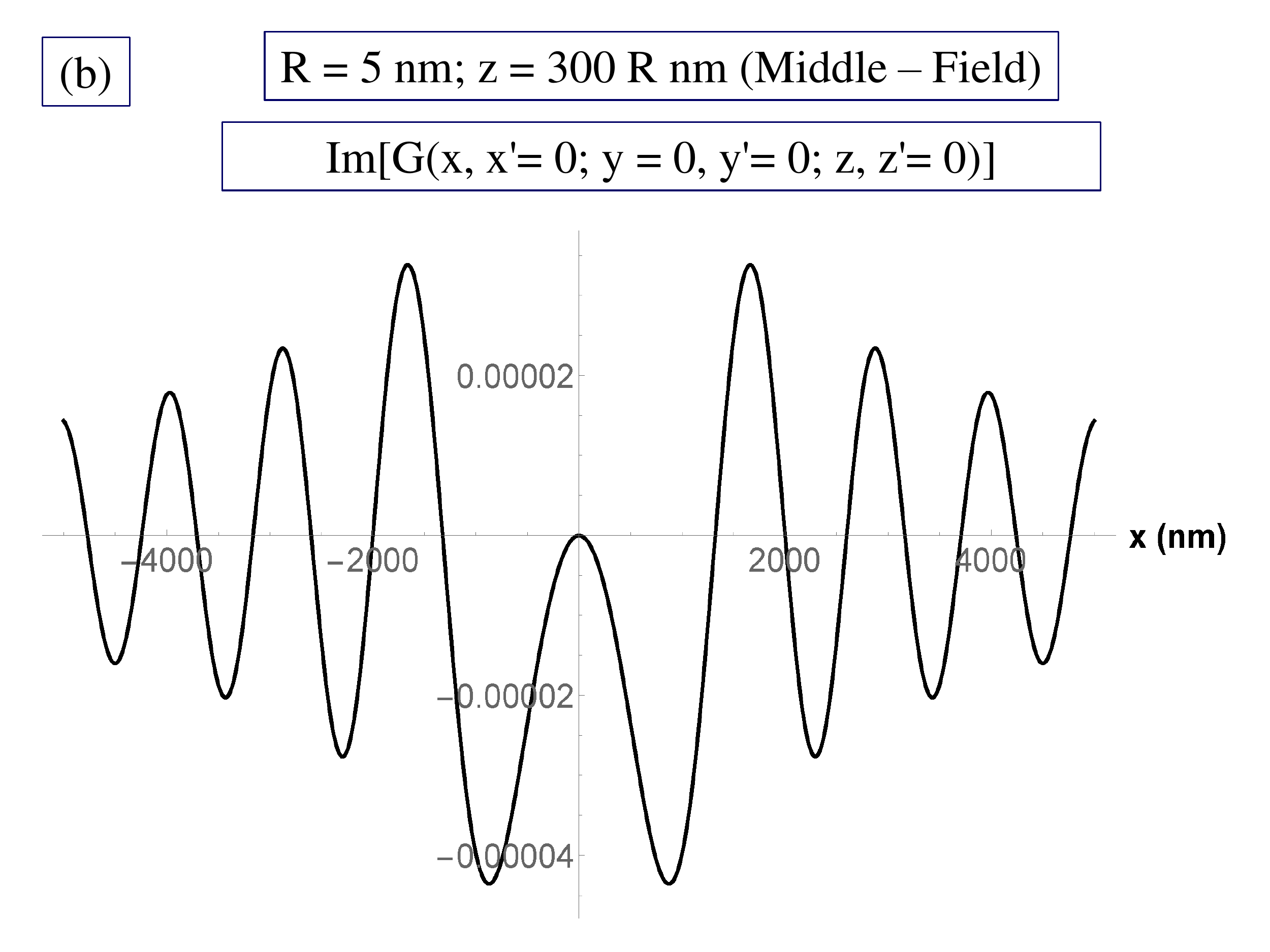}
\caption{$(a)$  exhibits Re$[G(x, x^{\prime}=0; y=0, y^{\prime}=0;z=300\,R, z^{\prime}=0;\omega)]$ and $(b)$ presents \\
Im $[G(x, x^{\prime}=0; y=0, y^{\prime}=0;z=300\,R, z^{\prime}=0;\omega)]$ as a function of $ x $ for a perforated 2D plasmonic layer of GaAs in the presence of a nano-hole of radius $ R=5\,nm $ at $ z = 300\,R $ (Middle-Field) for $ \varepsilon_{b}=1 $, $ n_{_{2D}}=4\times10^{15}cm^{-2} $, $ d=10\,nm $ and $ m^{\ast}=0.065 m_{0} $ where $ m_{0} $ is the free-electron mass.}
\label{MFR5GF}
\qquad
\end{figure*}

\begin{figure*}[htbp]
\centering
\includegraphics[width=8cm,height=7cm]{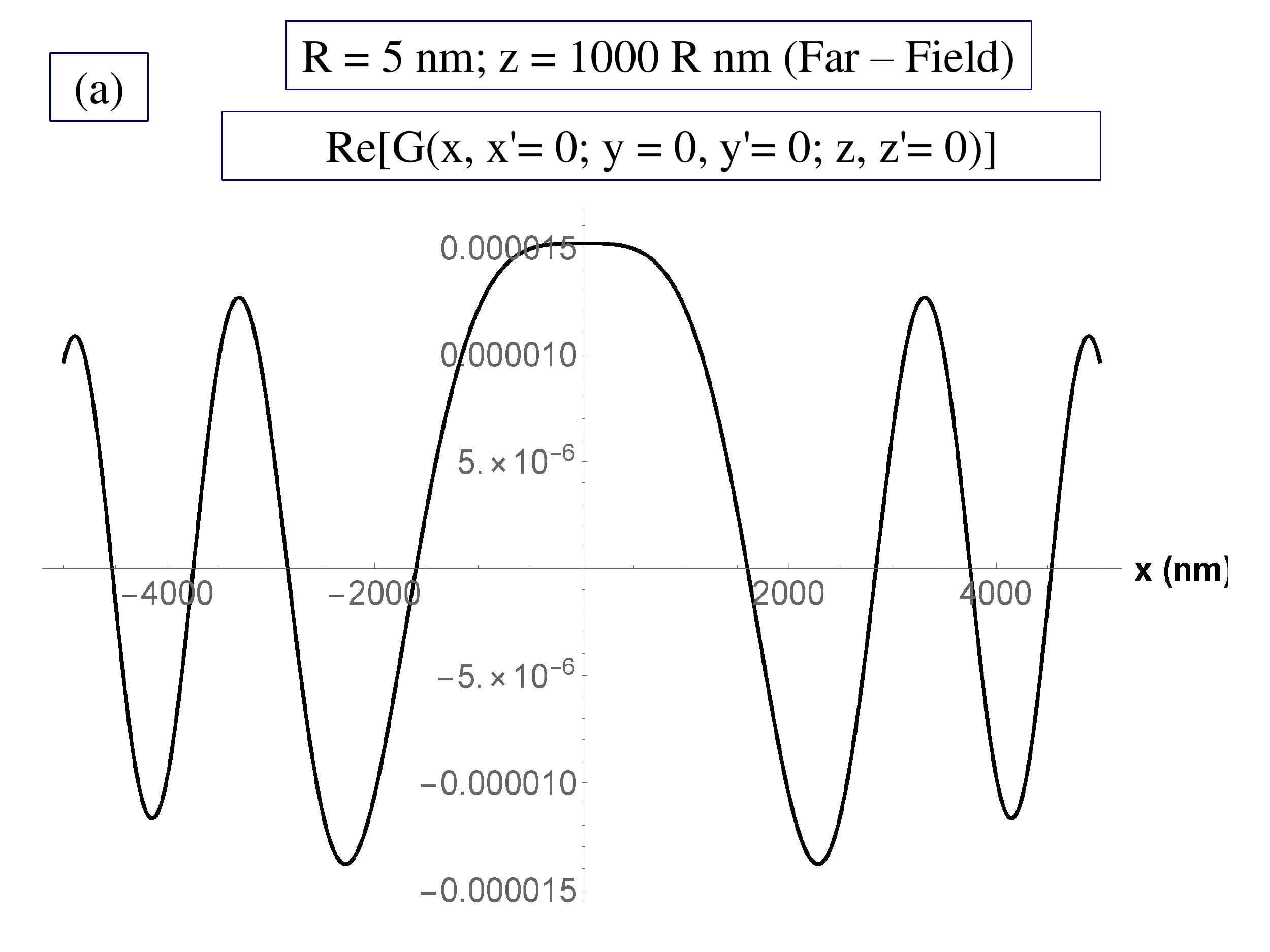}
\includegraphics[width=8cm,height=7cm]{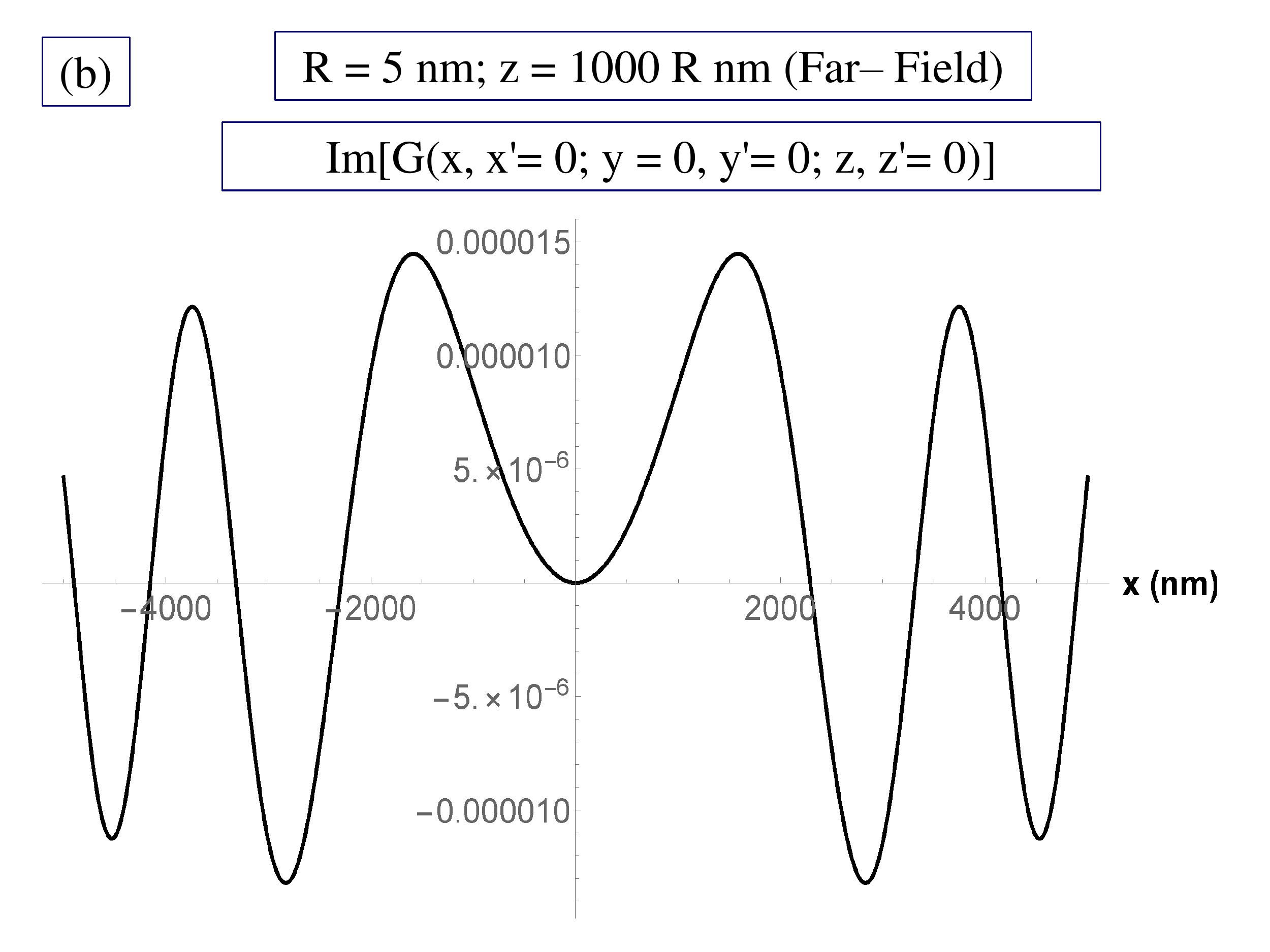}
\caption{$(a)$  exhibits Re$[G(x, x^{\prime}=0; y=0, y^{\prime}=0;z=1000\,R, z^{\prime}=0;\omega)]$ and $(b)$ presents \\
Im $[G(x, x^{\prime}=0; y=0, y^{\prime}=0;z=1000\,R, z^{\prime}=0;\omega)]$ as a function of $ x $ for a perforated 2D plasmonic layer of GaAs in the presence of a nano-hole of radius $ R=5\,nm $ at $ z = 1000\,R $ (Far-Field) for $ \varepsilon_{b}=1 $, $ n_{_{2D}}=4\times10^{15}cm^{-2} $, $ d=10\,nm $ and $ m^{\ast}=0.065 m_{0} $ where $ m_{0} $ is the free-electron mass.}
\label{FFR5GF}
\qquad
\end{figure*}
\newpage
\section{ Concluding Remarks}
\label{sec5}

In this paper we have carried out a thorough numerical analysis of the closed-form expression for
the scalar Green's function of a perforated, thin 2D plasmonic layer embedded in a 3D host medium in the presence of a nano-hole.

Inspection of the resulting Green's function figures shows that for large $ r_{\parallel} \mapsto x > 2500nm $ the spatial dependence of the Green's function becomes becomes oscillatory as a function of $ r_{\parallel} (x )$ with peaks uniformly spaced.  In this regard, it should be noted that our designation of near, middle and far radiation zones is defined in terms of $z$-values ($ 50R, 300R, 1000R $) $ alone $, to the exclusion of $ r_{\parallel}$:  In consequence of this exclusion, the figures actually carry useful information for $ r_{\parallel} (x )$ in $ all $ radiation zones as conventionally defined in terms of the incident wavelength $ \lambda \sim {2\,\pi}/{q_{\omega}}$.  Furthermore, this approach to oscillatory behavior as a function of $ r_{\parallel}$ with uniformly spaced peaks is accompanied by a geometric $1/ r_{\parallel}$ - diminution of the amplitude of the Green's function.  On the other hand, our far zone figures also show that when $ z > r_{\parallel}= x $ the Green's function flattens as a function of $ r_{\parallel}= x $ into a region of constancy, which is evident in Fig.\ref{FFR5GF}.

\numberwithin{equation}{section}
\numberwithin{figure}{subsection}


\section{Appendix }

We consider the evaluation of the following integral [7]

 \begin{equation}\label{Ap2.1}
I(a,b,c)=\int_{0}^{\infty} dx\ \frac{ x  J_{0}(ax)e^{{i}c\sqrt{b^{2}-x^{2}}}}{\sqrt{b^{2}-x^{2}}}=
\frac{e^{{i}b\sqrt{a^{2}+c^{2}}}}{i\sqrt{a^{2}+c^{2}}}.
 \end{equation}
 Setting $ c=0 $ in Eq.\  (\ref{Ap2.1})  yields

 \begin{equation}\label{Ap2.2}
I(a,b,0)=\int_{0}^{\infty}dx\ \frac{ x  J_{0}(ax)}{\sqrt{b^{2}-x^{2}}}=\frac{e^{{i}b\,a}}{i\,a}.
\end{equation}
Differentiating $ \partial^2/\partial{c}^2$ of Eq.\  (\ref{Ap2.1}), we obtain

\begin{eqnarray}\label{Ap2.3}
\frac{\partial^{2}\, I(a,b,c)}{\partial c^{2}}&=&\frac{\partial^{2}}{\partial c^{2}}\left\{
\int_{0}^{\infty}dx\ \frac{ x
 J_{0}(ax)e^{{i}c\sqrt{b^{2}-x^{2}}}}{\sqrt{b^{2}-x^{2}}}\right\}
 \nonumber\\
 &=&
\frac{\partial^{2}}{\partial c^{2}}\left\{
\frac{e^{{i}b\sqrt{a^{2}+c^{2}}}}{i\sqrt{a^{2}+c^{2}}}\right\}.
\nonumber\\
 \end{eqnarray}

\medskip
\par
Again setting $ c = 0 $  in Eq.\  (\ref{Ap2.3}) after taking the second derivative,
the following result is obtained

\begin{equation}\label{Ap2.4}
\int_{0}^{\infty}dx\  x \sqrt{(a^{2}-x^{2})}   J_{0}(x) = - (a+i)e^{ia}.
\end{equation}
The first derivative of Eq.\  (\ref{Ap2.1}) yields
\begin{eqnarray}\label{Ap2.5}
\frac{\partial\, I(a,b,c)}{\partial {c}}&=&\frac{\partial}{\partial {c}}\left\{
\int_{0}^{\infty}dx\ \frac{ x  J_{0}(ax)e^{{i}c\sqrt{b^{2}-x^{2}}}}{\sqrt{b^{2}-x^{2}}}\right\}=
\frac{\partial}{\partial{c}}\left\{
\frac{e^{{i}b\sqrt{a^{2}+c^{2}}}}{i\sqrt{a^{2}+c^{2}}}\right\}
\nonumber\\
&=&
\int_{0}^{\infty}dx\ { x  J_{0}(ax)e^{{i}c\sqrt{b^{2}-x^{2}}}}
=
-\,\frac{\partial}{\partial{c}}\left\{
\frac{e^{{i}b\sqrt{a^{2}+c^{2}}}}{\sqrt{a^{2}+c^{2}}}\right\}
\nonumber\\
&=&
-\,\frac{i c \left(i+\sqrt{a^2+c^2}\right) e^{i \sqrt{a^2+c^2}}}{\left(a^2+c^2\right)^{3/2}}.
\nonumber\\
\end{eqnarray}
\normalsize
\newpage
\begin{acknowledgments}
We gratefully acknowledge support from the NSF-AGEP program for the work reported in this paper.   Thanks are extended to Professors M. L. Glasser and Dr. Erik Lenzing for numerous helpful discussions as well as Liuba Zhemchuzhna and Dr. Andrii Iurov for helpful comments.
\end{acknowledgments}

\end{document}